\documentclass[10pt]{article}

\usepackage{amsmath,amssymb,amsfonts,amsthm,amsbsy}
\usepackage{mathtools}
\usepackage{multirow}
\usepackage{color}
\usepackage[a4paper,text={16.8cm,22.4cm}]{geometry}
\usepackage{graphicx}
\usepackage{caption}
\usepackage{slashed}
\usepackage{float}
\usepackage{subcaption}
\usepackage{appendix}
\usepackage{booktabs}
\usepackage{bm}
\usepackage{hyperref}
\usepackage{cite}
\usepackage{indentfirst}
\usepackage{simplewick}
\usepackage{orcidlink}
\usepackage[symbol]{footmisc}

\begin{document}
\begin{titlepage}

\begin{center}
{\bf\Large Determination of $B$-meson distribution amplitudes from $B\to \pi,K,D$ transition form factors}
\end{center}
\vspace{0.5cm}

\begin{center}
{\bf 
Dong-Hao Li$\orcidlink{0000-0002-6060-7926},^{a,}$
\footnote[1]{lidh@lzu.edu.cn}
Cai-Dian Lü$\orcidlink{0000-0002-5300-641X},^{b,c,}$
\footnote[2]{lucd@ihep.ac.cn}
Ulf-G.~Mei{\ss}ner$\orcidlink{0000-0003-1254-442X},^{d,e,f,}$
\footnote[3]{meissner@hiskp.uni-bonn.de}
and Jing Gao$\orcidlink{0000-0002-4892-9252}^{d,}$ 
\footnote[4]{corresponding author, gao@hiskp.uni-bonn.de}
}\\ 
\vspace{0.5cm}
{\sl $^a$\, MOE Frontiers Science Center for Rare Isotopes, \\and School of Nuclear Science and Technology, Lanzhou University, Lanzhou 730000, China}\\
{\sl $^b$ \,Institute of High Energy Physics, CAS, P.O. Box 918(4) Beijing 100049, China} \\
{\sl $^c$ \,School of Physics, University of Chinese Academy of Sciences, Beijing 100049, China} \\
{\sl $^d$ \,Helmholtz-Institut f\"{u}r Strahlen- und Kernphysik and Bethe Center for Theoretical Physics, Universit\"{a}t Bonn, D-53115 Bonn, Germany} \\
{\sl $^e$ \, Institute~for~Advanced~Simulation (IAS-4),~Forschungszentrum~J\"{u}lich,~D-52425~J\"{u}lich,~Germany}\\
{\sl $^f$ \, Peng Huanwu Collaborative Center for Research and Education, International Institute for Interdisciplinary and Frontiers, 
Beihang University, Beijing 100191, China}

\end{center}
\vspace{0.2cm}

\begin{abstract}
Recent work on $B \to \pi$, $K$ and $B\to D$ form factors from lattice QCD and light-cone sum rules has made it possible to constrain the inverse moment $\lambda_B$ of the $B$-meson light-cone distribution amplitudes by performing a global fit of $B\to \pi,K,D$ form factors.
We have compiled the $B\to \pi,K,D$ form factors calculated by the HPQCD, MILC, and RBC/UKQCD collaborations in the large $q^2$ region. 
By employing an three-parameter ansatz of the $B$-meson light-cone distribution amplitudes, we express the  $B\to \pi,K,D$ form factors at $q^2=0$ that are calculated from light-cone sum rules, in terms of the inverse moment $\lambda_B$ of the leading-twist $B$-meson light-cone distribution amplitude. 
In the $B \to \pi \ell \nu$ channel, we also include the available $q^2$-binned experimental data from the BaBar, Belle, and Belle~II collaborations.
Using the Bourrely-Caprini-Lellouch parametrization, we perform a global fit and obtain $\lambda_B=217(19)_{-17}^{+82}$ MeV and $|V_{\text{ub}}|=3.68(13)_{-1}^{+0}\times10^{-3}$. The second uncertainty is obtained by constraining $\lambda_B>200$ MeV and varying the inverse logarithmic moments $\hat{\sigma}_1\in[-0.7,0.7]$ and $\hat{\sigma}_2\in[-6,6]$, which represents the model-dependent uncertainty from the $B$-meson light-cone distribution amplitudes.
When taking into account $\lambda_B$ and $\hat{\sigma}_1$ as fitting parameters simultaneously, the intervals of our preditions are $\lambda_B=[208, 324]$ MeV and $\hat{\sigma}_1=[-0.7, 0.27]$. 

\end{abstract}

\vfil

\end{titlepage}

\section{Introduction}

The light-cone distribution amplitude (LCDA) is a fundamental object for establishing QCD factorization formula in exclusive hadronic processes \cite{Lepage:1979zb,Efremov:1979qk,Grozin:1996pq,Beneke:1999br,Keum:2003js,Hua:2020gnw,LatticeParton:2022zqc}. For $B$-meson semi-leptonic decays, the transition form factors can be calculated by lattice simulation at small recoil; while light-cone sum rules (LCSR) approach is the best theoretical method to calculate the transition form factors at large recoil with $B$-meson or final state meson LCDAs as input. Some studies exist of the pion LCDA, but have very limited knowledge of other meson LCDAs. Defined as light-like matrix element of heavy-quark effective theory (HQET) operators, $B$-meson LCDAs encode the momentum distribution of the light spectator parton at soft-scale and serve as an
indispensable ingredient for constructing light-cone sum rules of heavy-to-light hadronic matrix elements \cite{Khodjamirian:2005ea,Braun:2012kp}. However, in many cases the precision of theoretical predictions depends solely on the incomplete knowledge of $B$-meson LCDAs \cite{Cheng:2014fwa,Wang:2015vgv,Beneke:2020fot,Cui:2023jiw,Shen:2020hfq,Gao:2024vql,Huang:2024xii}, which also affects the precision of $|V_{ub}|$ or $|V_{cb}|$ determinations. 

Recently, significant progress has been made in understanding the model-independent properties or model-dependent parametrization of $B$-meson LCDAs, 
including its renormalization group equation as well as the underlying conformal symmetry \cite{Lange:2003ff,Bell:2013tfa,Braun:2014owa,Galda:2020epp,Huang:2023jdu,Feldmann:2022uok}, 
the perturbative QCD constraints based on the operator product expansion \cite{Lee:2005gza,Feldmann:2014ika},
the matching of the QCD LCDA and HQET LCDA \cite{Beneke:2023nmj},
model-independent determination of LCDA on the Euclidean lattice \cite{Wang:2019tgg,Wang:2019msf,Wang:2024wwa,LatticeParton:2024zko}, and the models for LCDAs which satisfy all existing constraints \cite{Braun:2017liq,Beneke:2018wjp}. 
A key parameter in modeling the LCDA is the inverse moment, $\lambda_B(\mu)$, of the leading-twist $B$-meson LCDA $\phi_B^+(w,\mu)$,
which is defined as $\lambda_B(\mu)^{-1}=\int_0^{\infty}\frac{dw}{w}\phi_B^+(w,\mu)$. 
Several complementary approaches have been developed to constrain this quantity. QCD sum rules provide a direct estimate of $\lambda_B$ invoking the quark-hadron duality ansatz, for instance, $\lambda_B(1\,\text{GeV})=460\pm110$~MeV \cite{Braun:2003wx} and $\lambda_B(1\,\text{GeV})=383\pm153$~MeV \cite{Khodjamirian:2020hob}. Experimental  measurements of the partial branching fractions of $B\to\gamma\ell\nu$ offer a lower bound $\lambda_B > 240$~MeV \cite{Belle:2018jqd,Wang:2016qii,Beneke:2018wjp}.
Preliminary results from lattice QCD, such as $\lambda_B=389(35)$~MeV, have started to provide insights into the nonperturbative properties of heavy-light meson distribution amplitudes, though they are still subject to significant systematic uncertainties \cite{LatticeParton:2024zko,Han:2024fkr}.
Indirect determinations can be made by comparing  $B\to \gamma$ form factors from LCSR predictions based on $B$-meson LCDAs with results from LCSR predictions based on photon distribution amplitudes, which yields $\lambda_B=360(110)$~MeV \cite{Janowski:2021yvz}. 
Similarly, the indirect extraction strategy can be applied to processes, such as $B\to \pi$, $B\to \rho$  and $B\to K$ form factors, leading to estimates of $\lambda_B$ in the range of approximately $300\sim 400$~MeV, albeit with significant model dependence as indicated by the spread of values obtained from different models of the $B$-meson LCDAs \cite{Wang:2015vgv,Gao:2019lta,Mandal:2023lhp}.

Although many discussions have been made to constrain the inverse moment in $B$-meson LCDAs, a global analysis using all available   $B$ to  pseudo-scalar meson transition form factors from lattice QCD calculation together with experimental measurements of semi-leptonic $B$ meson decays is missing. In this work, 
 we collect the lattice QCD results for $B\to\pi$, $B\to K$ and $B \to D$ transition form factors from the UKQCD collaboration \cite{Flynn:2015mha}, MILC collaboration \cite{FermilabLattice:2015mwy,Bailey:2015dka,FermilabLattice:2015ilb} and HPQCD collaboration \cite{Parrott:2022rgu,Na:2015kha} in the small recoil region.
Meanwhile, experimental data on the $q^2$-binned branching fractions of $B\to\pi\ell\nu$ from the BaBar, Belle and BelleII collaborations are also taken into account in our global analysis \cite{BaBar:2010efp,BaBar:2012thb,  Belle:2010hep,Belle:2013hlo,  Belle-II:2022imn}. In the large recoil region, we express these transition form factors in the LCSR approach with $B$-meson LCDAs, which have been achieved at next-to-leading-logarithmic (NLL) in the leading-power  and next-to-leading-power (NLP) in the leading-logarithmic  accuracy \cite{Cui:2022zwm, Gao:2021sav}.
For the $q^2$ dependence of the form factors, we use the Bourrely-Caprini-Lellouch (BCL) parameterization. Finally, a global fit of 
 both experimental and lattice QCD data 
  allows us to simultaneously extract the BCL parameters, the Cabibbo-Kobayashi-Maskawa (CKM) matrix element $|V_{\text{ub}}|$ and the inverse moment $\lambda_B(\mu)$ of the $B$-meson LCDA.

The organization of the article is as follows: In Section~2, we present the LCSR formulas of $B\rightarrow $ pseudoscalar meson transition form factors. In Section~3, we show the numerical details, including the fitting function, the resulting estimates of $|V_{\text{ub}}|$ and $\lambda_B$, and phenomenological analysis of the $B\to \pi\ell\nu$ observables. In Section~4, we draw our conclusions. 

\section{B to Pseudoscalar Meson Form Factors from Light-Cone Sum Rules}

The LCSR approach was first introduced in Refs.~\cite{Balitsky:1986st,Balitsky:1989ry} and later applied to exclusive $B$-meson decays in Ref.~\cite{Chernyak:1990ag}. In the  early stage, the LCSR approach was formulated using an $B$-meson interpolating current together with the light meson LCDAs \cite{hep-ph/9305348, hep-ph/9701238}. An alternative formulation of LCSR, which swap the meson state and interpolating current based on $B$-meson LCDAs, was later developed in Refs.~\cite{Khodjamirian:2005ea, Khodjamirian:2006st}. 
In this section, we introduce the theoretical framework of LCSR with $B$-meson LCDAs for $B$-meson to pseudoscalar transition form factors. 
For an overview of the current status and future prospects of LCSR, we refer the reader to Ref.~\cite{Khodjamirian:2023wol}. Further technical details on the calculation of QCD corrections, higher-twist contributions, and power corrections for $B\to \pi,K,D$ transition form factors can be found in Refs.~\cite{Wang:2015vgv,Wang:2017jow,Gubernari:2018wyi,Cui:2022zwm,Gao:2021sav}.

We begin with the vacuum-to-$B$-meson correlation function
\begin{equation}
\Pi_{\mu}(n\cdot p,\bar{n}\cdot p)=i\int \mathrm{d}^4 \mathrm{x}\,\mathrm{e}^{{i}p\cdot x}\langle 0|{T}\{\bar{d}(x)\slashed{n}\gamma_5 q_1(x),\bar{q}_1(0)\Gamma_\mu b(0)\}|\bar{B}(p_B)\rangle,
\label{eq-2pt}
\end{equation}
where $p$ is the momentum of the final pseudoscalar meson. In the interpolating current $\bar{d}\slashed{n}\gamma_5 q_1$, the quark field $q_1\in\{u,s,c\}$ corresponds to the pseudo-scalar meson $P\in\{\pi,K,D\}$, respectively. We have introduced two light-like vectors $n_\mu=(1,0,0,1)$ and $\bar{n}_\mu=(1,0,0,-1)$, and adopted the following power counting scheme,
\begin{equation}
    n\cdot p = \frac{m_B^2-q^2}{m_B}\sim \mathcal{O}(m_b),\qquad \bar{n}\cdot p \sim \mathcal{O}(\Lambda_{\text{QCD}}),
\end{equation}
where the transferred momentum is defined as $q\equiv p_B-p$, and the quadratic terms of mass $m_P$ are neglected at the leading power in the heavy quark expansion. The hadronic dispersion relation for the correlation function $\Pi_\mu$ can be obtained by inserting a complete set of hadron states in Eq.~(\ref{eq-2pt}),
\begin{equation}
    \begin{aligned}
       \Pi_{\mu}(n\cdot p,\bar{n}\cdot p)=\frac{\langle 0|\bar{d}(x)\slashed{n}\gamma_5 q_1(x)|P(p)\rangle\langle P(p)|\bar{q}_1\Gamma_{\mu}b|\bar{B}(p_B)\rangle}{m^2_P-p^2}    +\frac{1}{2\pi}\int_{s_0^h}^{\infty}\mathrm{ds}\frac{\rho_\mu(s)}{s-p^2},
    \end{aligned}\label{eq-2pt-hadron}
\end{equation}
where $\rho_\mu(s)$ represents the contributions from the higher excited and continuum states. The vacuum-to-$P$-meson matrix element in Eq.~(\ref{eq-2pt-hadron}) is proportional to the decay constant, $\langle 0|\bar{d}\slashed{n}\gamma_5 q_1|P(p)\rangle=\mathrm{i}n\cdot p \,f_P$.
With the different Dirac structures of weak currents, $\Gamma_\mu=\gamma_\mu, \sigma_{\mu\nu}q^\nu$, the vector form factor $f_{BP}^+(q^2)$,  the scalar form factor $f_{BP}^0(q^2)$, and tensor form factor $f_{BP}^T(q^2)$ can be extracted from the first term in Eq.~(\ref{eq-2pt-hadron}), which are defined by
\begin{equation}
    \begin{aligned}
    &\langle P(p)|\bar{q}_1 \gamma_\mu b|\bar{B}(p_B) \rangle 
    = 
    f^+_{BP}(q^2) 
    \left[ p_B + p - \frac{m_B^2 - m_P^2}{q^2} q \right]_\mu + f^0_{BP}(q^2) \frac{m_B^2 - m_P^2}{q^2} q_\mu , 
    \\ 
    &\langle P(p)|\bar{q}_1 \sigma_{\mu\nu} q^\nu b|\bar{B}(p_B) \rangle 
    = 
    \frac{\mathrm{i} f^T_{BP}(q^2)}{m_B + m_P} 
    \left[ q^2 (2p + q)_\mu - (m_B^2 - m_P^2) q_\mu \right].
\end{aligned}\label{eq-formfactor-defination}
\end{equation}
To construct the sum rules for the $B\to \pi,K,D$ transition form factors, two essential components are required: the light-cone operator product expansion, and the dispersion relation with the assumption of quark-hadron duality.
Subsequently, we take the  derivation of tree level form factor $f_{B\pi}^+(q^2)$ as an example. For the external momenta far below the hadronic thresholds,
    $p^2\ll 0$ and $ q^2\ll m_B^2$,
the operator product expansion near the light-cone $x^2\simeq 0$ can be applied to the correlation function in Eq.~(\ref{eq-2pt}). The resulting factorization expression is a convolution of the perturbative kernel with the $B$-meson LCDA,
\begin{equation}
    \begin{aligned} 
    \Pi_{\mu}^{\text{tree}}(n \cdot p, \bar{n} \cdot p) & =
    {i} \int \mathrm{d}^{4} \mathrm{x} \,\mathrm{e}^{{i} p \cdot x}\langle 0| \bar{d}(x) \slashed{n} \gamma_{5} \contraction{}{u}{(x)}{\bar{u}} u(x) \bar{u}(0) \gamma_{\mu} b(0)\left|\bar{B}(p_B)\right\rangle \\ & =
    -{i} f_{B} m_{B} \bar{n}_{\mu} \int_{0}^{\infty} \mathrm{d \omega} \frac{\phi_{B}^{-}(\omega, \mu)}{\bar{n} \cdot p-\omega+{i} 0}.
    \end{aligned}
    \label{eq-2pt-treelevel}
\end{equation}
Employing the dispersion relation 
\begin{equation}
    \Pi_{\mu}(n \cdot p, \bar{n} \cdot p)=\frac{1}{\pi}\int_{0}^{\infty}\mathrm{d \omega^{\prime}} \frac{\operatorname{Im}_{\omega^{\prime}} \Pi_{\mu}\left(n \cdot p, \omega^{\prime}\right)}{\omega^{\prime}-\bar{n} \cdot p-{i} 0},
\end{equation}
the quark-hadron duality assumption above an effective threshold $w_s$,
\begin{equation}
    \frac{1}{2\pi}\int_{s_{0}^h/n\cdot p}^{\infty} \mathrm{d \omega^{\prime}}\frac{\rho_{\mu}\left( \omega^{\prime}\right)}{\omega^{\prime}-\bar{n} \cdot p-{i} 0} 
    =
    \frac{1}{\pi}\int_{\omega_{s}}^{\infty}\mathrm{d \omega^{\prime}} \frac{\operatorname{Im}_{\omega^{\prime}} \Pi_{\mu}\left(n \cdot p, \omega^{\prime}\right)}{\omega^{\prime}-\bar{n} \cdot p-{i} 0} ,
\end{equation}
and applying the Borel transformation $\bar{n}\cdot p \to w_M$, the sum rule for $f_{B\pi}^+(q^2)$ can be derived by matching the partonic representation of the correlation function in Eq.~(\ref{eq-2pt-treelevel}) with its hadronic representation in Eq.~(\ref{eq-2pt-hadron}),
\begin{equation}
    f_{B\pi}^{+}\left(q^{2}\right)=
    \frac{f_{B} m_{B}}{f_{\pi} n \cdot p} \operatorname{Exp}\left[\frac{m_{\pi}^{2}}{n \cdot p \, w_{M}}\right] \int_{0}^{w_{s}} \mathrm{d \omega^{\prime}} \operatorname{Exp}\left(-\frac{\omega^{\prime}}{w_{M}}\right) \phi_{B}^{-}(\omega,\mu),
\end{equation}
where the Borel parameter $w_M$ and the effective threshold $w_s$ are two intrinsic parameters in the LCSR approach. 
The $B$-meson LCDA $\phi_B^-(\omega,\mu)$ is defined through a non-local matrix element \cite{Bell:2013tfa}, 
\begin{equation}
    \begin{aligned}
        \langle 0| \bar{d}_{\beta}(\tau \bar{n})[\tau \bar{n}, 0] b_{\alpha}(0)|\bar{B}(p_B)\rangle 
        =-\frac{{i} f_{B} m_{B}}{4}\left\{\frac{1+\slashed{v}}{2}\left[2 \tilde{\phi}_{B}^{+}(\tau)+\left(\tilde{\phi}_{B}^{-}(\tau)-\tilde{\phi}_{B}^{+}(\tau)\right) \slashed{n}\right] \gamma_{5}\right\}_{\alpha \beta}
    \end{aligned}
\end{equation}
with the Fourier transformations,
\begin{equation}
    \phi_{B}^{ \pm}(\omega^{\prime},\mu)=\int_{-\infty}^{+\infty} \frac{\mathrm{d \tau}}{2 \pi} e^{{i} \omega^{\prime} \tau} \tilde{\phi}_{B}^{ \pm}(\tau,\mu),\quad
    \tilde{\phi}_{B}^{ \pm}(\tau,\mu)
    =\int_{0}^{+\infty}\mathrm{d \omega'}
    e^{-{i} \omega^{\prime} \tau}\phi_{B}^{ \pm}(\omega^{\prime},\mu),
\end{equation}
and   velocity $v$ of the $B$-meson satisfying $n\cdot v=\bar{n}\cdot v =1$ and $v_\perp=0$. The (logarithmic) inverse moments of the leading-twist $B$-meson LCDA $\phi_B^+(\omega,\mu)$ are defined by
\begin{equation}
    \begin{aligned}
    \frac{1}{\lambda_{B}(\mu)}
    =
    \int_{0}^{\infty} \frac{\mathrm{d \omega}}{\omega} \phi_{B}^{+}(\omega, \mu), \quad 
    \frac{\widehat{\sigma}_{n}(\mu)}{\lambda_{B}(\mu)}
    =
    \int_{0}^{\infty} \frac{\mathrm{d \omega}}{\omega} \ln ^{n} \frac{e^{-\gamma_{E}} 
    \lambda_{B}(\mu)}{\omega} \phi_{B}^{+}(\omega, \mu).
    \end{aligned}
\end{equation}
Currently, both experimental and theoretical constraints on the $B$-meson LCDAs are still insufficient. We therefore adopt a three-parameter ansatz for the $B$-meson LCDAs \cite{Beneke:2018wjp}, which satisfies the known constrains, including the equations of motion for different twists and the asymptotic behavior at large scales. Under this ansatz, the LCDA is expressed as hypergeometric function $U(a, b, z)$   
\begin{equation}
    \phi_B^{+}(\omega)=\frac{\Gamma(\beta)}{\Gamma(\alpha)} \frac{\omega}{\omega_{0}^{2}} e^{-\omega / \omega_{0}} U\left(\beta-\alpha, 3-\alpha, \omega / \omega_{0}\right),
\end{equation}
where the parameters $\omega_0,\alpha$ and $\beta$ in this model can be determined via 
\begin{equation}
    \begin{aligned} 
    \lambda_{B}(\mu) & =\frac{\alpha-1}{\beta-1} \omega_{0}, 
    \\ 
    \widehat{\sigma}_{1}(\mu) & =\psi(\beta-1)-\psi(\alpha-1)+\ln \frac{\alpha-1}{\beta-1}, 
    \\ 
    \widehat{\sigma}_{2}(\mu) & =\widehat{\sigma}_{1}^{2}(\mu)+\psi^{\prime}(\alpha-1)-\psi^{\prime}(\beta-1)+\frac{\pi^{2}}{6},
    \end{aligned}\label{eq:3para-3inversemoment}
\end{equation}
with  $\psi(x)$   the digamma function.
Analytical expressions and the leading-logarithmic evolution for the LCDAs are collected in the appendix of our previous works \cite{Gao:2021sav,Gao:2024vql}. 

In this work, we adopt the following ranges of these parameters:
\begin{equation}
\lambda_B \in \left[200, 500\right]~\text{MeV}, \quad \hat{\sigma}_1 \in \left[-0.7, 0.7\right], \quad \hat{\sigma}_2 \in \left[-6, 6\right],
\end{equation}
at the renormalization scale $\mu_0=1$ GeV. 
The interval $\lambda_B \in \left[200, 500\right]~\text{MeV}$ is recommended from the indirect determination in $B\to\gamma\ell\bar{\nu}$ channel \cite{Wang:2016qii,Wang:2018wfj,Janowski:2021yvz}, which encompasses most of the theoretical prescriptions. The ranges for $\hat{\sigma}_1$ and $\hat{\sigma}_2$ reflect the uncertainty arising from the model dependence of the $B$-meson LCDA, due to our limited knowledge of its precise form.

\section{Numerical Results}

\begin{table}[htbp]
\renewcommand{\arraystretch}{1.5}
\centering
\caption{The value of   input parameters   in our numerical analysis.}
\begin{tabular}{|l|l|l||l|l|l|}
\hline
Parameter & Value & Ref. & Parameter & Value & Ref. \\
\hline
\hline
$\overline{m}_b(\overline{m}_b)$ & $4.200(14)~\mathrm{GeV}$  & \cite{FlavourLatticeAveragingGroupFLAG:2024oxs}  & $\bar{\Lambda}$ & $0.48(10)~\mathrm{GeV}$ & \cite{Cui:2022zwm} \\
$m_{B}$ & $5.27966 ~\mathrm{GeV}$ & \cite{ParticleDataGroup:2024cfk} & $m_{B^*(1-)}$ & $5.3247 ~\mathrm{GeV}$ & \cite{ParticleDataGroup:2024cfk} \\
$m_{B_c^*(1-)}$ & $6.332 ~\mathrm{GeV}$ & \cite{Dowdall:2012ab} & $m_{B_s^*(1-)}$ & $5.4154~\mathrm{GeV}$ & \cite{ParticleDataGroup:2024cfk} \\
$m_{B_c^*(0+)}$ & $6.712 ~\mathrm{GeV}$ & \cite{Mathur:2018epb} & $m_{B_s^*(0+)}$ & $5.718 ~\mathrm{GeV}$ & \cite{Bardeen:2003kt} \\
\hline
\hline
& & & & $\{0.7,6.0\}$ & \\
$\lambda_B(\mu_0)$& $350(150)~\mathrm{MeV}$ &\cite{Beneke:2020fot} & $\{ \hat{\sigma}_1(\mu_0), \hat{\sigma}_2(\mu_0) \}$ & $\{0.0,\pi^2/6\}$ & \cite{Beneke:2020fot}\\
& & & & $\{-0.7,-6.0\}$ & \\
\hline
\end{tabular}
\label{tab:parameters}
\end{table}

The LCSR calculations of the semileptonic $B$-meson decay form factors have reached NLL and NLP accuracy \cite{Gao:2021sav,Cui:2022zwm,Cui:2023jiw}. 
However, the precision of these theoretical predictions is currently limited by the systematic uncertainties in $B$-meson LCDAs. 
Despite considerable efforts have made to constrain $B$-meson LCDAs, our understanding remains rather limited, even for its first inverse moment, $\lambda_B$. 
In this section, we propose a complementary approach to determine $\lambda_B$: a global fit to the $B\to \pi,K,D$ form factors. This strategy broadly follows that of conventional fits used to extract the CKM matrix elements $|V_{ub}|$ or $|V_{cb}|$, with one essential difference: the $\lambda_B$-dependent form factors derived from LCSR are not treated as input data, but as parameters to be determined from the global fit. 

\begin{figure}[htb]
\centering
\includegraphics[width=0.8\textwidth]{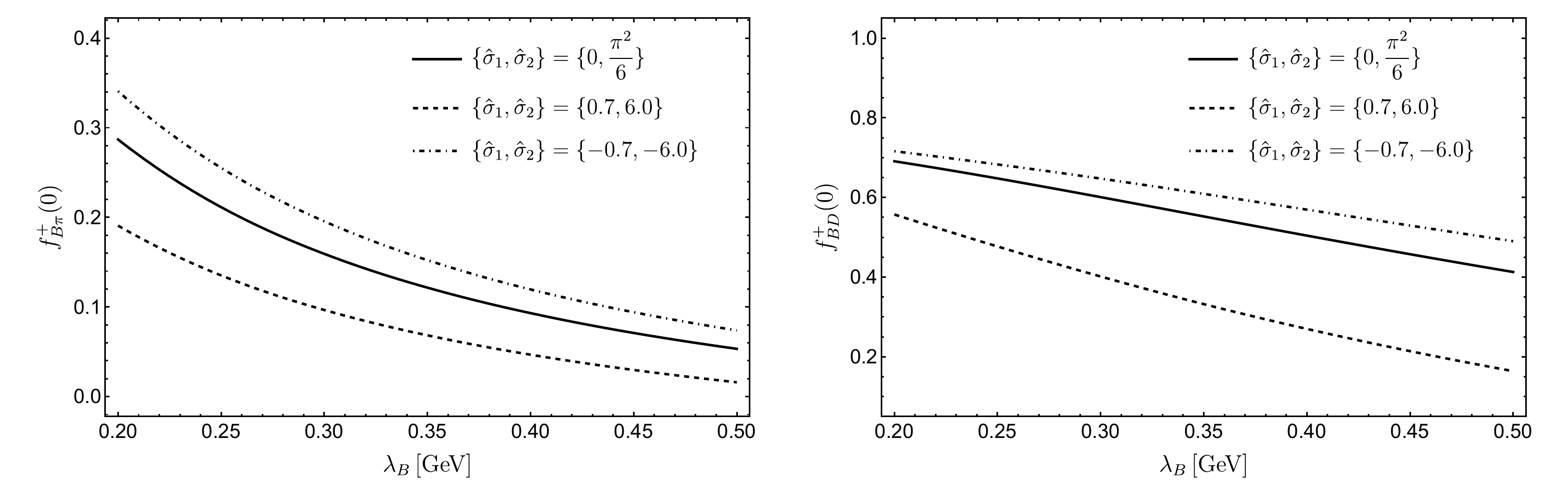}
\includegraphics[width=0.8\textwidth]{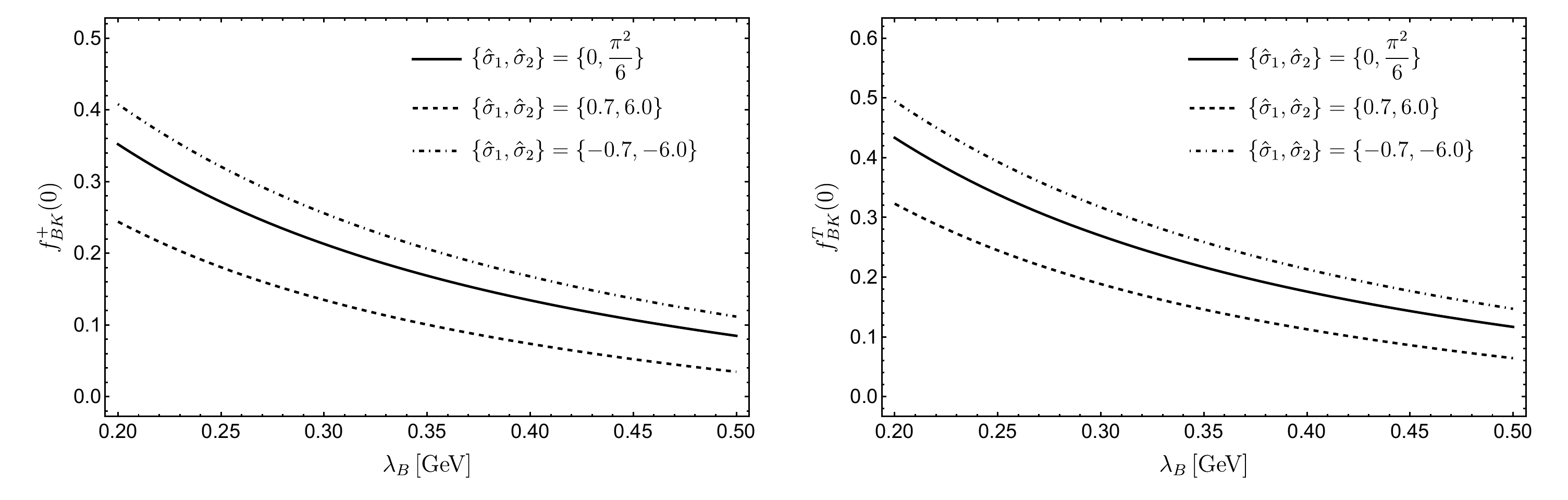}
\caption{The dependence of the $B \to P$ form factors on $\lambda_B$ for three different sets of $\{\hat{\sigma}_1, \hat{\sigma}_2\}$ . For the $B \to \pi,\,K$ cases, the Borel parameter is $M^2 = 1.25$~GeV$^2$, and the effective threshold parameters are $s_0^\pi = 0.7$~GeV$^2$ and $s_0^K = 1.05$~GeV$^2$. For the $B \to D$ case, the Borel parameter $M^2 = 4.5$~GeV$^2$, and the effective threshold parameter is $s_0^D = 6.0$~GeV$^2$.}
\label{fig:fflambdaB}
\end{figure}

 For input parameters required in the calculation of $B $ to pseudoscalar form factors with LCSR, we adopt   values from Refs.~\cite{Gao:2021sav, Cui:2022zwm}, as summarized in Tab.~\ref{tab:parameters}. In particular, we use the bottom-quark mass $\bar{m}_b(\bar{m}_b) = 4.200(14)$~GeV in the Modified Minimal Subtraction scheme and the effective mass $\bar{\Lambda}=m_B-m_b^{\mathrm{pole}}$ with the pole mass $m_b^{\mathrm{pole}}=4.8(1)$~GeV. We then illustrate the dependence of the $B\to \pi,K,D$ form factors on the inverse moment $\lambda_B$ for three distinct parameter sets in Fig.~\ref{fig:fflambdaB}. Due to the different values for LCSR parameters, $M^2, s_0$, used in $B\to D$ and $B\to \pi, K$ processes, the dependence of the $B\to D$ form factors on the inverse moment $\lambda_B$ exhibits a different behavior from that in other processes.

We adopt the BCL parametrization \cite{Bourrely:2008za} for the $q^2$ dependence of the $B $ to pseudoscalar meson form factors, which has been widely employed in $B$ decays. 
The form factors $f_{BP}^+(q^2)$ are then parameterized as  
\begin{equation}
    f_{BP}^{+}\left(q^{2}\right)=\frac{1}{1-q^{2} / m_{B_{q}^{*}}^{2}} \sum_{i=0}^{N-1} b_{P,i}^{+}\left[z\left(q^{2}, t_{0}\right)^{i}-(-1)^{i-N} \frac{i}{N} z\left(q^{2}, t_{0}\right)^{N}\right],
    \label{eq:BCLf+}
\end{equation}
where the masses of the $J^P=1^-$ resonant mesons $m_{B_q^*}$ are taken as $m_{B_d^*}\,(m_{B_s^*},\,m_{B_c^*})$ for $f_{B\pi}^{+}\,(f_{BK}^{+},\,f_{BD}^{+})$.
The conformal map $z(q^2,t_0)$ is defined as
\begin{equation}
z(q^2,t_0) \equiv \frac{\sqrt{t_+ - q^2} - \sqrt{t_+ - t_0}}{\sqrt{t_+ - q^2} + \sqrt{t_+ - t_0}},
\label{eq:zmap}
\end{equation}
with $t_{\pm}=(m_B\pm m_P)^2$. The parameter $t_0$ is taken as $t_{0}=t_{+}-\sqrt{t_{+}\left(t_{+}-t_{-}\right)}$, which can maximally reduce the range of $z$ after mapping the domain $q^2\in\left[ 0,(m_B-m_P)^2 \right]$. We adopt the truncation order $N=3$ in the $z$-series expansion, which is sufficient to satisfy the current lattice QCD constraints on the form factors \cite{Cui:2022zwm}. Along the same lines, the scalar form factor $f_{BP}^0(q^2)$ and the tensor form factor $f_{BK}^T(q^2)$ can be parametrized as
\begin{equation}
\begin{aligned}
  &f_{B\pi}^{0}\left(q^{2}\right)=\sum_{i=0}^{N-1} b_{\pi,i}^{0} z\left(q^{2}, t_{0}\right)^{i},\quad 
  f_{B K}^{0}\left(q^{2}\right)=\frac{1}{1-q^{2} / m_{B_{s}^{*}(0+)}^{2}} \sum_{i=0}^{N-1} b_{K,i}^{0} z\left(q^{2}, t_{0}\right)^{i},\\
  & 
  f_{B D}^{0}\left(q^{2}\right)=\frac{1}{1-q^{2} / m_{B_{c}^{*}(0+)}^{2}} \sum_{i=0}^{N-1} b_{D,i}^{0} z\left(q^{2}, t_{0}\right)^{i},
  \\
  &
  f_{BK}^{T}\left(q^{2}\right)=\frac{1}{1-q^{2} / m_{B_{s}^{*}}^{2}} \sum_{i=0}^{N-1} b_{K,i}^{T}\left[z\left(q^{2}, t_{0}\right)^{i}-(-1)^{i-N} \frac{i}{N} z\left(q^{2}, t_{0}\right)^{N}\right],\\
\end{aligned}\label{eq:BCLf0fT}
\end{equation}
with the predicted masses $m_{B_s^*(0+)}=5.718$~GeV and $m_{B_c^*(0+)}=6.712$ GeV from Refs.~\cite{Bardeen:2003kt,Mathur:2018epb}. It is noted that alternative parameterizations for the $B\to D$ form factors are available, especially the Boyd–Grinstein–Lebed parametrization, which readily accommodates the strong unitarity bound~\cite{Boyd:1994tt,Boyd:1995sq,Boyd:1997kz}. 
For further discussion of the unitarity bounds see \cite{Gambino:2019sif,DiCarlo:2021dzg,Martinelli:2021onb,Jaiswal:2020wer,Leljak:2021vte}.
However, as the strong unitarity bound requires a comprehensive treatment of all $b\to c$ exclusive channels, which is beyond the scope of this work, we ultimately employ the BCL parameterization for our numerical calculations.

In the small recoil region, lattice QCD provides high-precision calculations of the $B\to \pi,K,D$ form factors. To be specific, the lattice data used in this work include: 11 data points for $f_{B\pi}^{+,0}(q^2)$ from the RBC/UKQCD and FNAL/MILC collaborations \cite{FermilabLattice:2015mwy,Flynn:2015mha}; 16 data points for $f_{BK}^{+,0,T}(q^2)$ from the HPQCD and FNAL/MILC collaborations \cite{Bailey:2015dka,Parrott:2022rgu}; and 12 data points for $f_{BD}^{+,0}(q^2)$ from the HPQCD and FNAL/MILC collaborations\cite{FermilabLattice:2015ilb,Na:2015kha}.
It is convenient to define the $\chi^2$-function,
\begin{equation}
\chi^2_{\text{LQCD}} = \sum_{P} \sum_{i,j} \left[ f_{BP}(q^2_i; \vec{b}) - f_{BP,i}^{\text{LQCD}} \right] \left(\mathrm{Cov}_{\mathrm{lat}}^{-1} \right)_{ij} \left[ f_{BP}(q^2_j; \vec{b}) - f_{BP,j}^{\text{LQCD}} \right],
\label{eq:LQCDchi2}
\end{equation}
where $f_{BP}(q^2_i; \vec{b})$ denote the form factor expressed in terms of the BCL parameters $\vec{b}$, $f_{BP,i}^{\text{LQCD}}$ denote the values of the corresponding lattice QCD results at the points $q_i^2$, and Cov$_{\mathrm{lat}}$ is the covariance matrix among the lattice data points.
In Refs.~\cite{FermilabLattice:2015mwy,Bailey:2015dka,Parrott:2022rgu}, the lattice QCD collaborations did not provide specific data points, instead, they provided the results of their BCL fit to the data points. Adopting the method from Ref.~\cite{Cui:2022zwm}, we use the lattice fit results to generate correlated data points in large $q^2$ region.
The numerical values of the lattice data are listed in Appendix~\ref{app:data}.

In the large-recoil region, the definition of the LCSR part of $\chi^2$-function differs slightly from the conventional one used in $|V_{ub}|$ extractions:
\begin{equation}
\chi^2_{\text{LCSR}} = \sum_{m,n=1}^4 \left[ f_m(0;\vec{b}) - f_m^{\text{LCSR}}(0;\lambda_B) \right] \left( \mathrm{Cov}^{-1} \right)_{mn} \left[ f_n(0;\vec{b}) - f_n^{\text{LCSR}}(0;\lambda_B) \right],
\label{eq:chi2LCSR}
\end{equation}
where $f_m(0;\vec{b})\in\left\{f_{B\pi}^+(0),f_{BK}^+(0),f_{BK}^T(0),f_{BD}^+(0)\right\}$ are the BCL parametrization  form factors at $q^2=0$, and $f_m^{\text{LCSR}}(0;\lambda_B)$ denotes the LCSR       form factors as a function of the inverse moment $\lambda_B$, which is displayed in Fig.~\ref{fig:fflambdaB}. Specifically, $f_m^{\text{LCSR}}(0;\lambda_B)$ is treated as a parameter to be fitted, rather than as the input data. 
The covariance matrix $\mathrm{Cov}$ for LCSR results in Eq.~(\ref{eq:chi2LCSR}) is composed of two components:
\begin{equation}
\mathrm{Cov} = (0.1 f_{m}^{\text{LCSR}}) (\mathrm{Cov}_{\text{uncor}})_{ mn} (0.1 f_{n}^{\text{LCSR}}) + (0.1 f_{m}^{\text{LCSR}}) (\mathrm{Cov}_{\text{cor}})_{ mn} (0.1 f_{n}^{\text{LCSR}}),
\label{eq:cov_LCSR}
\end{equation}
where the first term, $(0.1 f_{m}^{\text{LCSR}}) (\mathrm{Cov}_{\text{uncor}})_{ m,n} (0.1 f_{n}^{\text{LCSR}})$, represents uncorrelated systematic uncertainties originating from the effective threshold $s_0$ and Borel parameter $M_2$ in the LCSR calculation. These uncertainties are treated as uncorrelated among the different form factors $f_{B\pi}, f_{BK}$ and $f_{BD}$.
The second term, $(0.1 f_{m}^{\text{LCSR}}) (\mathrm{Cov}_{\text{cor}})_{ m,n} (0.1 f_{n}^{\text{LCSR}})$, accounts for fully correlated ($100\%$ correlated) statistical uncertainties arising from the remaining input parameters in the LCSR framework, such as the factorization scale and HQET parameters.

We then define the $\chi^2$-function for the experimental data as
\begin{equation}
    \chi_{\text {exp }}^{2}=\sum_{j, k=1}^{57}\left[\mathcal{B}_{j}^{\text {exp}}-\mathcal{B}_{j}\left(f^{+}_{B\pi}\right)\right] (\mathrm{Cov}_{\mathrm{exp}}^{-1})_{jk}\left[\mathcal{B}_{k}^{\text {exp }}-\mathcal{B}_{k}\left(f^{+}_{B\pi}\right)\right],
    \label{eq:chi2-exp}
\end{equation}
where $\mathcal{B}_{j}^{\text {exp}}$ denotes the experimentally measured $q^2$-binned partial branching fraction of $B\to\pi\ell\bar{\nu}_{\ell}$ and $\mathrm{Cov}_{\mathrm{exp}}$ is the covariance matrix of the corresponding experimental datasets from the BaBar \cite{BaBar:2010efp,BaBar:2012thb},Belle \cite{Belle:2010hep,Belle:2013hlo} and Belle II \cite{Belle-II:2022imn} Collaborations. The partial branching fraction $\mathcal{B}_{j}\left(f^{+}_{B\pi}\right)$ is obtained by integrating
\begin{equation}
    \frac{d \Gamma}{d q^{2}}\left(B \rightarrow \pi \ell \bar{\nu}_{\ell}\right)=\frac{G_{F}^{2}\left|V_{u b}\right|^{2}}{192 \pi^{3} m_{B}^{3}} \left[
    (m_B^2+m_\pi^2-q^2)^2-4m_B^2m_\pi^2
    \right]^{3 / 2}\left|f_{+}\left(q^{2}\right)\right|^{2}
    \label{eq:BR-ff}
\end{equation}
over the corresponding $q^2$ interval, neglecting the mass of the lepton.
It is worth noting that the LCSR approach based on $B$-meson LCDAs features a unique advantage: the $B$-meson LCDAs, which serve as non-perturbative and universal inputs, enter in all semileptonic $B$ decay form factors. 
By incorporating the high-precision lattice QCD results for the $B\to K $ and $B\to D$ form factors, the global fit can significantly improve the accuracy of $B\rightarrow\pi$ form factor $f_{B\pi}^+(0)$, which ultimately contributes to a more reliable extraction of CKM matrix element $|V_{ub}|$.

\begin{figure}[!ht]
\centering
\includegraphics[width=0.95\textwidth]{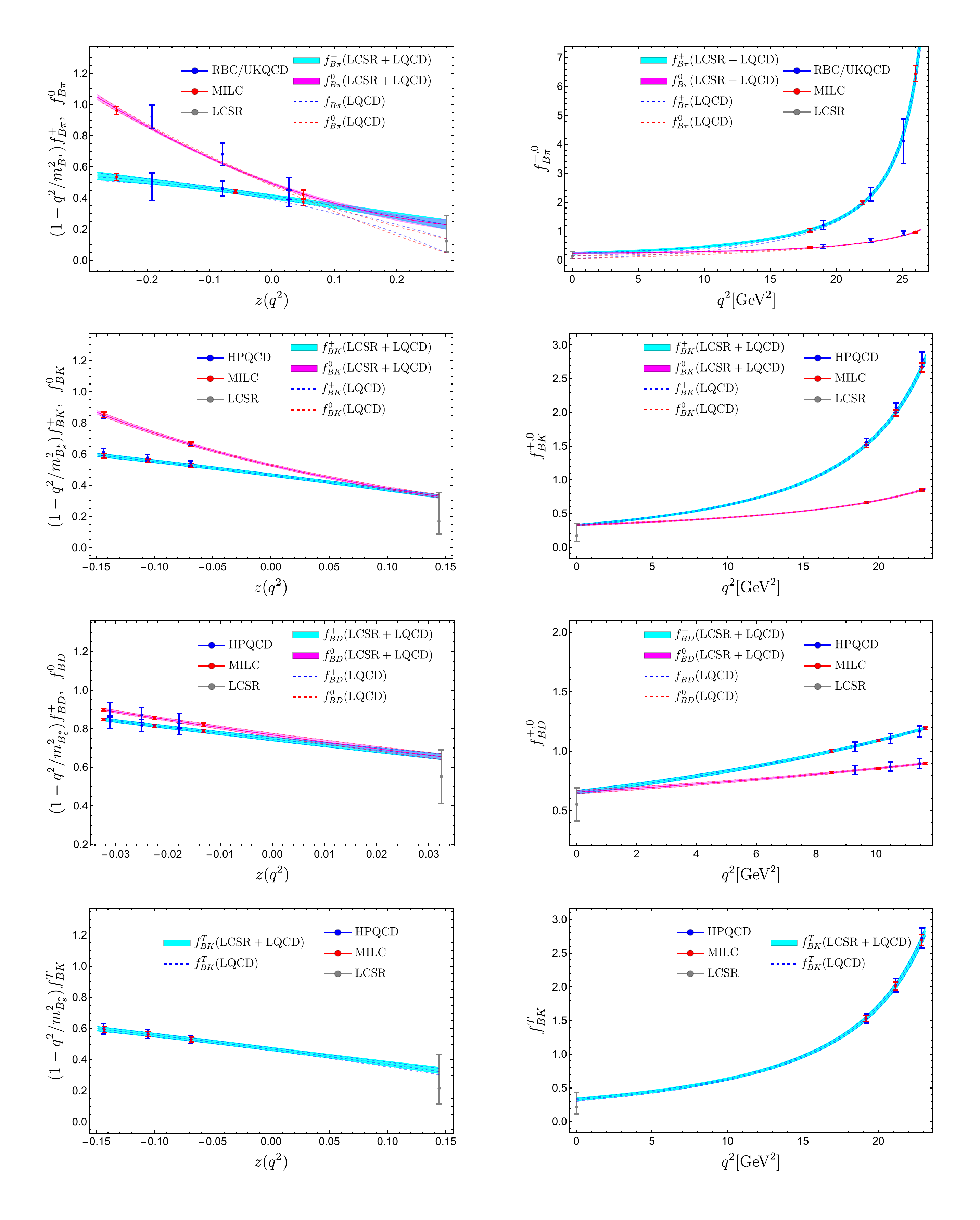}
\vspace{-20pt}
\caption{Results of the global fit to the $B \to P$ form factors versus $z$ (left pannel) and versus $q^2$ (right pannel) with the parameter set $\{\hat{\sigma}_1, \hat{\sigma}_2\} = \{0, \pi^2/6\}$. The gray points correspond to the LCSR form factor at $q^2=0$ for $\lambda_B = 350~\text{MeV}$, with the upper and lower limits of the gray error bars representing for $\lambda_B = 200~\text{MeV}$ and $\lambda_B = 500~\text{MeV}$, respectively. We also display the results by performing a BCL fit to input data only from lattice QCD as dashed curves for a comparison.}
\label{fig:ff-fit}
\end{figure}

\begin{figure}[htb]
\centering
\includegraphics[width=0.8\textwidth]{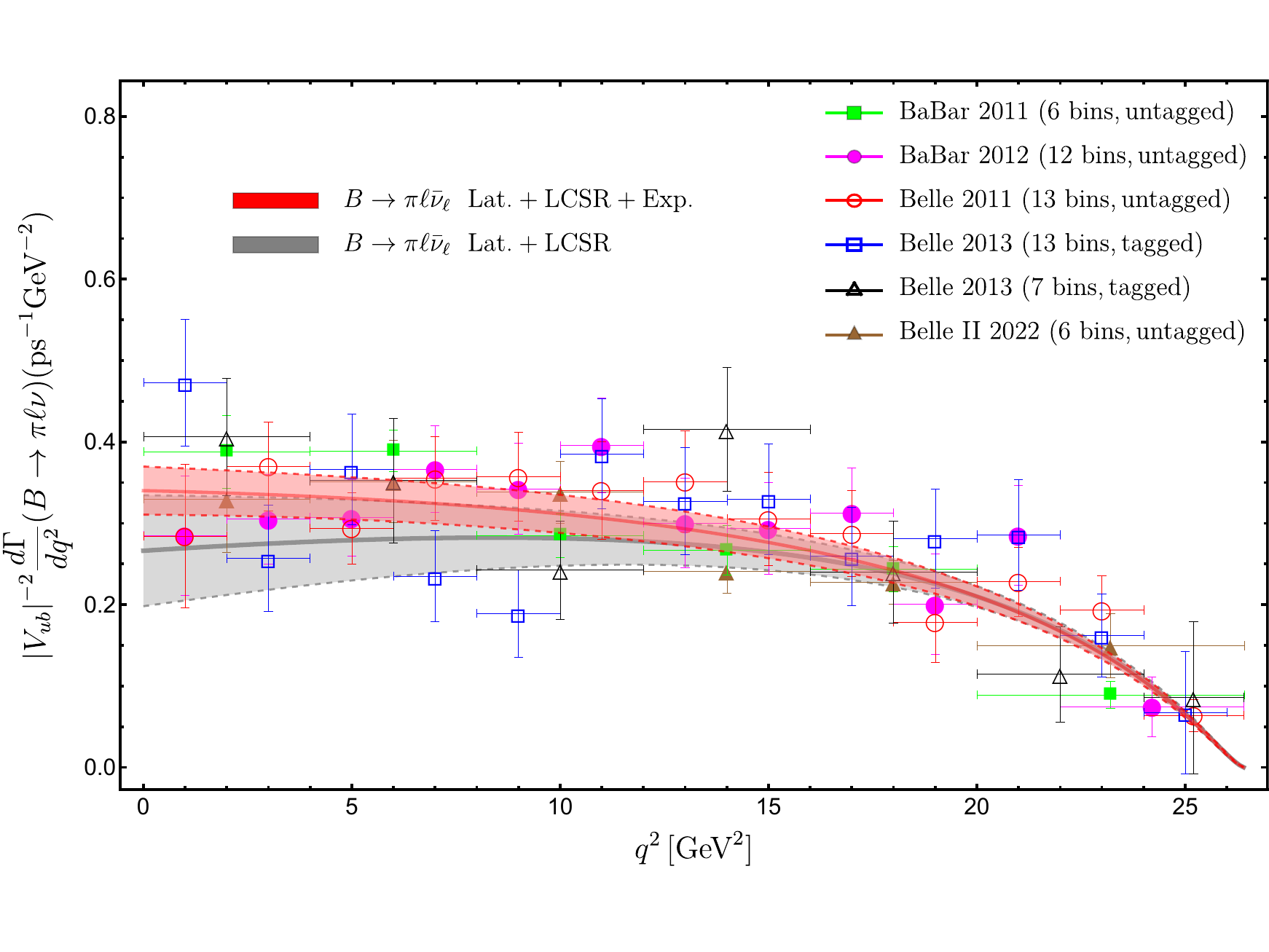}
\vspace{-20pt}
\caption{Theoretical predictions for the CKM-independent differential $q^2$ distributions of the $B \to \pi \ell \bar{\nu}_\ell$ process, using the $z$-series parameters obtained from the global fit.}
\label{fig:expvub}
\end{figure}

We subsequently perform a global fit to the above inputs by minimizing the total $\chi^2$ function,
\begin{equation}\label{chi2tot}
\chi^2(\lambda_B,|V_{ub}|,\vec{b})=\chi^2_{\text{LQCD}}+\chi^2_{\text{LCSR}}+\chi^2_{\text{exp}},
\end{equation}
using the BCL parametrization formulas given in Eqs.~(\ref{eq:BCLf+}) and (\ref{eq:BCLf0fT}) with the truncation order $N=3$. The global fit has 20 free parameters: $\lambda_B$, $|V_{ub}|$ and 18 BCL coefficients $\vec{b}$ with 96 data points. The number of degrees of freedom (d.o.f) is $96 - 20=76$, and we obtain $\chi^2_{\text{min}}/ \mathrm{dof} = 84.7/76 \approx 1.1 $. The fitting results of the $B\rightarrow P$ form factors in the  entire kinematic region and CKM-independent $q^2$-differential branching fractions of $B\to\pi\ell\bar{\nu}_{\ell}$ are shown in Figs.~\ref{fig:ff-fit} and \ref{fig:expvub}, respectively.

The numerical results of the global fit are shown as follows,
\begin{equation}\label{eq:fitresult}
    \lambda_B = 217(19)_{-17}^{+82}\,\text{MeV},\qquad |V_{ub}| = 3.68(13)_{-1}^{+0} \times 10^{-3},  
\end{equation}
and
\begin{equation}
\setlength{\arraycolsep}{1pt}
\begin{aligned}
b_{\pi,+}^0 &= 0.408(12),   & b_{\pi,+}^1 &= -0.509(47),  & b_{\pi,+}^2 &= -0.12(17), & b_{\pi,0}^0 &= 0.495(8),   & b_{\pi,0}^1 &= -1.411(37), \\
b_{K,+}^0   &= 0.466(8),    & b_{K,+}^1   &= -0.902(55),   & b_{K,+}^2   &= -0.28(26), & b_{K,0}^0   &= 0.291(4),   & b_{K,0}^1   &= 0.260(39), \\
b_{D,+}^0   &= 0.747(9),    & b_{D,+}^1   &= -2.97(18),    & b_{D,+}^2   &= 4.0(2.3),  & b_{D,0}^0   &= 0.662(8),   & b_{D,0}^1   &= -0.17(19), \\
b_{K,T}^0   &= 0.470(10),    & b_{K,T}^1   &= -0.899(82),  & b_{K,T}^2   &= -0.24(36).
\end{aligned}
\label{eq:fitresult2}
\end{equation}
The central values in Eqs.~(\ref{eq:fitresult}) and (\ref{eq:fitresult2}) are obtained under the default setting $\hat{\sigma}_1=0$, $\hat{\sigma}_2=\pi^2/6$ and with the constraint $\lambda_B >200$~MeV. The first uncertainty (shown in parentheses) represents the statistical error from the input parameters, while the second uncertainty corresponds to the systematic error obtained by varying the parameters $\hat{\sigma}_1\in [-0.7,0.7]$ and $\hat{\sigma}_2\in [-6,6]$. The systematic uncertainties on the BCL coefficients are not listed, as they are negligible and have no substantial impact. 
It is worth emphasizing that the LCSR approach based on light meson LCDA provides direct predictions for the $B\to \pi,K$ form factors at zero momentum transfer $q^{2}=0$, and that several parameterizations together with unitarity bounds for $B\to \pi,K$ form factors are available in Refs. \cite{Lellouch:1995yv,Becher:2005bg,Duplancic:2008ix,Bharucha:2010im,Bharucha:2012wy,Khodjamirian:2011ub,Gonzalez-Solis:2021pyh,SentitemsuImsong:2014plu,Khodjamirian:2017fxg,Gambino:2020jvv,Leljak:2021vte,EOSAuthors:2021xpv,Flynn:2023qmi,Gubernari:2023puw}. The above results are obtained from a global fit to experimental measurements and lattice QCD inputs, while the light meson LCSR computed form factors are not included as additional constraints in our fit. A quantitative assessment for the impact of incorporating light meson LCSR data is presented in Appendix~\ref{app:ann}.
For the $B\to K$ and $B\to D$ channels, the $z$ series expansion converges rapidly because the semileptonic region $0\le q^{2}\le (m_{B}-m_{K,D})^{2}$ is well separated from the nearest singularities associated with the branch cut. The lattice QCD data are sufficiently precise to allow reliable extrapolations to $q^{2}=0$, thereby enabling accurate determinations of $f_{BK}(0)$ and $f_{BD}(0)$ with a moderate truncation order \cite{Martinelli:2021onb,Martinelli:2021frl,Gubernari:2023puw}. 
For the $B\to \pi$ channel, the convergence of the $z$-expansion is less favorable, and extrapolating lattice QCD results to the low $q^2$ region can be problematic. We therefore include the experimentally measured $q^2$-binned branching fractions in our fit. In this setup, the global normalization quantity $|V_{ub}|$ is primarily determined by the precise form factors at several discrete $q^{2}$ points from lattice, while the $q^{2}$ dependence of $f_{B\pi}^{+}(q^{2})$ is constrained by the experimental data, thereby enabling a reliable determination of $f_{B\pi}^{+}(0)$ and of $\lambda_{B}$.
We stress that, for precise theoretical predictions of the lepton-flavour universality ratio $R_{\pi}$ and angular observables such as the forward-backward asymmetry $A_{\rm FB}$ and flat term $F_H$, the LCSR computed form factors in the small $q^{2}$ region, together with the unitarity bounds are indispensable. Nevertheless, since our analysis relies on a combined fit incorporating a large set of experimental data, the impact of including additional light meson LCSR inputs on our final results is expected to be relatively limited.

In Fig.~\ref{fig:ff-fit}, the dashed curves represent the $B\to \pi,K,D$ form factors computed by lattice QCD, while the color bands show the global fitting results of the form factors. It can be seen that the lattice input data in the $B\to K$ and $B\to D$ channels are quiet precise, so the global fit has little impact on these two channels. 
In contrast, the global fit improves the precision of the $B\to \pi$ form factor in the low $q^2$ region. In Fig.~\ref{fig:expvub}, we display the experimentally measured partial branching fractions, as well as the theoretical predictions for the differential branching fractions obtained from global fit. 
The resulting CKM matrix element $|V_{ub}|=3.67(13)\times10^{-3}$ is in good agreement with $|V_{ub}|=3.71(13)\times10^{-3}$ obtained from the global fit to $B\to \pi$ and $B\to D$ form factors in Ref.~\cite{Cui:2023jiw}, as well as $ V_{ub} = 3.70(10)(12)\times 10^{-3}$ extracted from exclusive decays in Particle Data Group \cite{ParticleDataGroup:2024cfk}.
To estimate the systematic uncertainty associated with the truncated order $N$ in the BCL parameterization, we repeated the fit with truncation order $N=4$. The resulting values of $\lambda_B$ and $|V_{ub}|$ differ from those obtained with $N=3$ by about $(1-1.5)\sigma$, confirming the stability of our conclusions.

\begin{figure}[ht]
\centering
\includegraphics[width=0.45\textwidth]{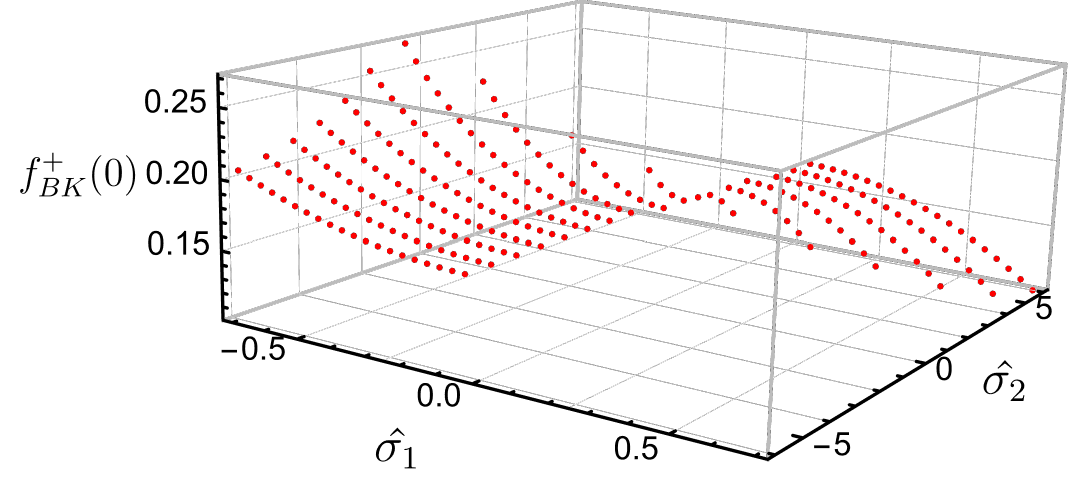}
\hspace{15pt}
\includegraphics[width=0.4\textwidth]{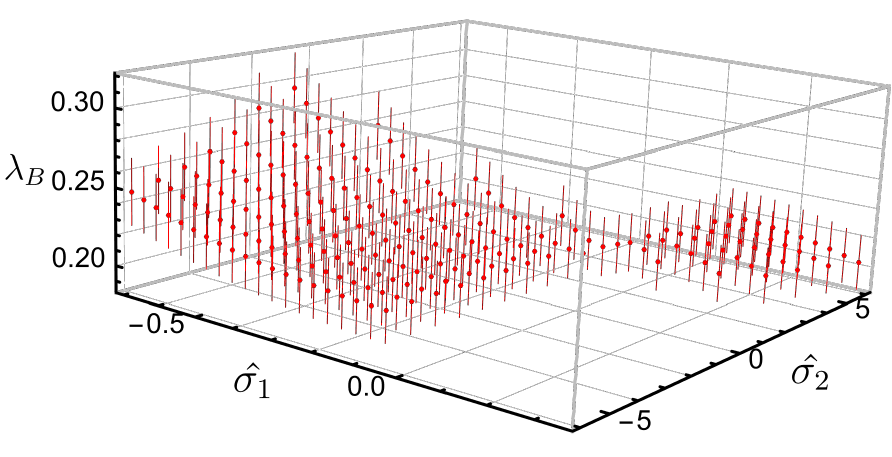}
\caption{Dependence of the form factor $f_{BK}^+(0)$ (left panel) and the best-fit value of $\lambda_B$ (right panel) on the parameters $\hat{\sigma}_1$ and $\hat{\sigma}_2$.}
\label{fig:3dBK}
\end{figure}

Taking the $B\to K$ decay channel as example, we show the dependence of the form factor $f_{BK}^+(0)$ on the parameters $\{\hat{\sigma}_1,\hat{\sigma}_2\}$ and present the corresponding best-fit values of $\lambda_B$ for different choices of $\{\hat{\sigma}_1,\hat{\sigma}_2\}$ in Fig.~\ref{fig:3dBK}. We can find that $\lambda_B$ shows a clear correlation with $\hat{\sigma}_1$: smaller values of $\hat{\sigma}_1$ correspond to larger values of $\lambda_B$. In addition, it makes sense that the parameters $\{\hat{\sigma}_1,\hat{\sigma}_2\}$ cannot take arbitrary values within their conventional ranges, because they are constrained by the three-parameter model. Only the sets $\{\hat{\sigma}_1,\hat{\sigma}_2\}$ located within the region bounded by the Eq.~(\ref{eq:3para-3inversemoment}) can be inverted to model parameters $\alpha$ and $\beta$.

It can be seen that the determination of the inverse moment, $\lambda_B=217(19)_{-17}^{+82}$~MeV, is still subject to sizable uncertainties due to the model dependence of $B$-meson LCDAs. However, this has no impact on the form factors, such as $f_{B\pi}^+(0)=0.258(11)$, and the resulting CKM matrix element $|V_{ub}|=3.67(13)\times 10^{-3}$.
Moreover, within the current theoretical framework, imposing the constraint $\lambda_B>200$~MeV in the global fit implies $\hat{\sigma}_1<0.35$, while $\hat{\sigma}_2$ remains essentially unconstrained. This reflects the fact that the form factors are more sensitive to $\hat{\sigma}_1$ than $\hat{\sigma}_2$.
In Tab.~\ref{tab:values of lambdaB}, we summarize results of the inverse moment $\lambda_B$ obtained from various theoretical methods. The central value of our result is lower than all those listed in the table, because the LCSR inputs incorporate NLP corrections to the form factors. This NLP correction decreases the magnitude of the form factors by approximately $30\%$, consequently leading to a smaller extracted value of $\lambda_B$. We also perform the global fit using LCSR form factors computed at NLL accuracy without NLP contributions, this yields $\lambda_B = 321(26)^{+84}_{-48}\,\text{MeV}$, which is in good agreement with the majority of the theoretical predictions shown in Tab.~\ref{tab:values of lambdaB}.

\begin{figure}[ht!]
\centering
\vspace{-30pt}
\includegraphics[width=0.9\textwidth]{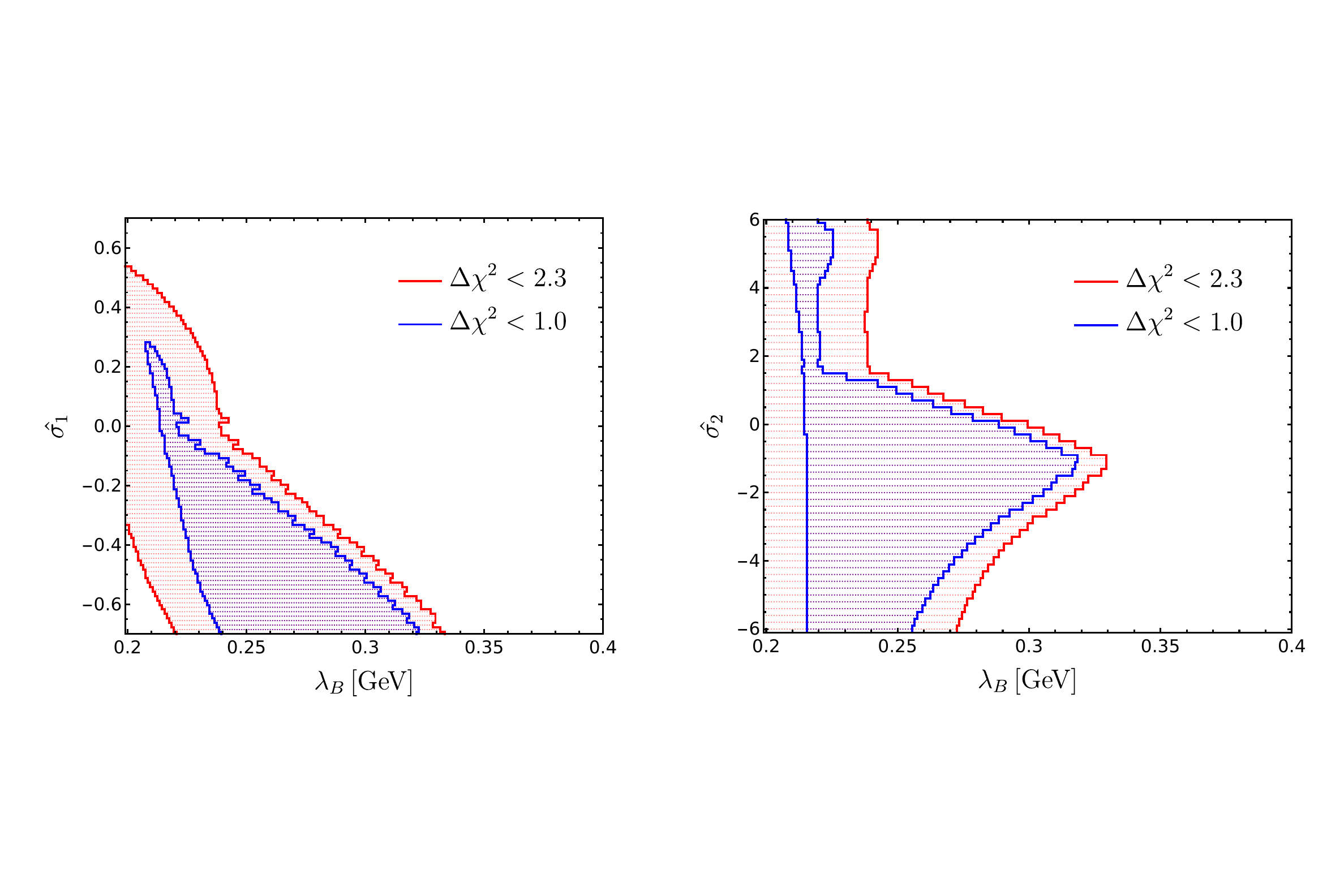}
\vspace{-60pt}
\caption{The joint $68.3\%$ confidence region (red contour) for the parameters $\{\lambda_B, \hat{\sigma}_1\}$ (left panel) and $\{\lambda_B, \hat{\sigma}_2\}$ (right panel). The projections of the blue region onto the $\lambda_B$, $\hat{\sigma}_1$ and $\hat{\sigma}_2$ axes yield the individual $68.3\%$ confidence intervals for each parameter.}
\label{fig:2dcov}
\end{figure}

Since the dependence of the $B\rightarrow P$ form factors on both $\lambda_B$ and $\hat{\sigma}_1$ is significant, one should perform a simultaneous analysis of both $\lambda_B$ and $\hat{\sigma}_1$.
We therefore define
\begin{equation}
\Delta\chi^2 = \chi^2_{\min}(\lambda_B,\hat{\sigma}_1) - \chi^2_{\min},
\end{equation}
where $\chi^2_{\min} = 82.5$ is the global minimum in Eq.~(\ref{chi2tot}), and $\chi^2_{\min}(\lambda_B,\hat{\sigma}_1)$ denotes the minimal $\chi^2$ value  for a fixed set $\{\lambda_B,\hat{\sigma}_1\}$,
with the free parameters including the BCL coefficients $\vec{b}$, $|V_{ub}|$ and $\hat{\sigma}_2$.
By scanning the $\{\lambda_B,\hat{\sigma}_1\}$ parameter space and repeatedly minimizing $\chi^2(\lambda_B,\hat{\sigma}_1)$ at each point, we can construct contours of $\Delta\chi^2=\text{constant}$, which correspond to joint confidence regions for the parameters $\{\lambda_B,\hat{\sigma}_1\}$.
The individual $68.3\%$ confidence intervals for $\lambda_B$ and $\hat{\sigma}_1$ are obtained by selecting the region where $\Delta\chi^2 \leq 1$, 
\begin{equation}
    \lambda_B = [ 208 , 324] \, \mathrm{MeV}, \qquad \hat{\sigma}_1 = [-0.7 , 0.27].
    \label{eq:fit-2dresult}
\end{equation}
It should be noted that we do not use complicated analytical expressions to find the minimum $\chi^2_{\min}(\lambda_B,\hat{\sigma}_1)$ for arbitrary parameter sets. Instead, we discretize $\lambda_B$ and $\hat{\sigma}_1$ and perform a uniform grid scan to evaluate $\Delta\chi^2$. The resolution of the scan is $1\,\mathrm{MeV}$ in $\lambda_B$ and approximately $0.01$ in $\hat{\sigma}_1$.
In this way, we display the joint $68.3\%$ confidence region for $\{\lambda_B, \hat{\sigma}_1\}$ in Fig.~\ref{fig:2dcov}, which is defined by the contour $\Delta\chi^2 = 2.3$. 
Similarly, we can also obtain the confidence interval for $\hat{\sigma}_2$, but the $\hat{\sigma}_2$ is not constrained by the joint fit due to the insensitivity to the form factors.
Consequently, we list this global fitting result in Tab.~\ref{tab:values of lambdaB} as well, in comparison with other theoretical predictions \cite{Wang:2016qii,Mandal:2023lhp,Gao:2019lta,Janowski:2021yvz,Wang:2015vgv}.
Obviously, the theoretical uncertainty is getting smaller after the global fit, especially for the statistical errors. 

\begin{table}[htbp]
\renewcommand{\arraystretch}{1.5}
\centering
\caption{A summary of theoretical determinations of $\lambda_B$ at the renormalization scale $\mu_0 = 1\,\mathrm{GeV}$. The two values from this work are given in Eqs.~(\ref{eq:fitresult}) and (\ref{eq:fit-2dresult}). Multiple results from the same reference arise from the different models of $B$-meson LCDA.}
\begin{tabular}{|c|c|c||c|c|c|}
\hline
Approach &  $\lambda_B~(\mathrm{MeV})$ & Ref &  
Approach &  $\lambda_B~(\mathrm{MeV})$ & Ref \\
\hline
\hline
\multirow{2}{*}{QCD sum rule } &  $460(110)$ & \cite{Braun:2003wx} &  \multirow{2}{3cm}{The upper limit of $\mathcal{BR}(B\to\gamma\ell\bar{\nu}_{\ell})$ }  &  $>240$ & \cite{Belle:2018jqd} \\
  &  $383(153)$ & \cite{Khodjamirian:2020hob} & 
  &  $>214$ & \cite{Wang:2016qii} \\
  \hline
  lattice QCD & $376(63)$ & \cite{Han:2024fkr,LatticeParton:2024zko}
  & \multirow{2}{3cm}{$B\to K$ form factor} & $338_{-9}^{+68}$ & \cite{Mandal:2023lhp}\\
  \cline{1-3}
$B\to \gamma$ form factor & $360(110)$ & \cite{Janowski:2021yvz} &
 & $472_{-41}^{+110}$ & \cite{Mandal:2023lhp}
  \\
\hline
\multirow{2}{3cm}{$B\to \rho$ form factor} &$370_{-86}^{+69}$ & \cite{Gao:2019lta} &
\multirow{4}{3cm}{$B\to \pi$ form factor} &$354_{-30}^{+38}$ & \cite{Wang:2015vgv}\\
& $343_{-64}^{+79}$ &  \cite{Gao:2019lta}&  
& $368_{-32}^{+42}$ &   \cite{Wang:2015vgv} \\
\cline{1-3}
\multirow{2}{3cm}{Global fit to $B\to \pi,K,D$ form factor} &$217(19)_{-17}^{+82}$ & This work &
& $389_{-28}^{+35}$ &    \cite{Wang:2015vgv} \\
 &$[208,324]$ & This work &
 & $303_{-26}^{+35}$& \cite{Wang:2015vgv} \\
\hline
\end{tabular}
\label{tab:values of lambdaB}
\end{table}

\section{Conclusions}

In this work, we investigated the dependence of the $B \to  \pi, K, D$ form factors on the inverse moment $\lambda_B$, using light-cone sum rules with the three-parameter model of the $B$-meson LCDAs. In contrast to   previous studies extracting $\lambda_B$ by comparing LCSR predictions with only some lattice QCD form factors, we adopted a complementary strategy: performing a global fit to all available $B \to  \pi, K, D$  form factors, including the lattice QCD results from the RBC/UKQCD, FNAL/MILC, and HPQCD collaborations, as well as the partial branching fractions of $B \to \pi \ell \bar{\nu}_\ell$ decay measured by the BaBar, Belle, and Belle~II experiments. We obtain
\[
\lambda_B = 217(19)^{+82}_{-17}~\text{MeV}, \qquad |V_{\text{ub}}| = 3.67(13)^{+0}_{-1} \times 10^{-3},
\]
where the systematic uncertainties arising from the $B$-meson LCDAs model are estimated by imposing the constraint $\lambda_B>200$ MeV and varying the inverse logarithmic moments within the intervals $\hat{\sigma}_1 \in [-0.7,\, 0.7]$ and $\sigma_2 \in [-6,\, 6]$. Although the systematic error on $\lambda_B$ remains sizable, the extracted CKM matrix element $|V_{\text{ub}}|$ is precisely constrained by the lattice QCD inputs and is insensitive to the $B$-meson LCDAs modeling uncertainty. Given the strong correlation between $\lambda_B$ and $\hat{\sigma}_1$—both of which have a substantial impact on the form factors. We further performed a profile $\chi^2$ fit to determine their joint confidence region. The individual $68.3\%$ confidence intervals for $\lambda_B$ and $\hat{\sigma}_1$ are obtained by the region satisfying $\Delta\chi^2 \leq 1$, 
\begin{equation*}
    \lambda_B = [ 208 , 324] \, \mathrm{MeV}, \qquad \hat{\sigma}_1 = [-0.7 , 0.27].
\end{equation*}

Future improvements of this analysis can be pursued along several directions. 
First, our current LCSR calculation is carried out at leading power contributions up to $\mathcal{O}(\alpha_s)$ and NLP corrections at $\mathcal{O}(\alpha_s^0)$ accuracy. A theoretical upgrade including two- and three-particle higher-twist B-meson LCDAs contributions at NLL in QCD would further enhance the precision of the form factors, and thereby increase the accuracy of  the $\lambda_B$ determination. 
Second, incorporating the experimental data from $b \to c$ semileptonic decays together with strong unitarity constraints could simultaneously determine $|V_{cb}|$ and the ratio $|V_{ub}/V_{cb}|$. 
Third, significant uncertainties arise in the extrapolation of the BCL-parametrized form factor from the small-recoil to the large-recoil region. Including multiple LCSR-calculated form factor points near $q^2 \approx 0$ can substantially provide a better description of the form factors in the entire kinematic region and increase the precision of $\lambda_B$ extraction.
Finally, the intrinsic LCSR parameters, such as the Borel mass $M^2$ and effective threshold $s_0$ should in principle depend on $\lambda_B$. Including $M^2$ and $s_0$ as free parameters in the global fit, or establishing their dependence on $\lambda_B$, would allow a more realistic evaluation of LCSR-related systematic uncertainties and lead to a more robust global analysis.

\section*{Acknowledgement}
We thank Yu-Ming Wang for valuable discussions. 
We thank Yong-Kang Huang and Yan-Bing Wei for providing the code to calculate the $B\to \pi,K,D$ form factors in LCSR.
This work is partly  supported  by the National Key Research and Development Program of China
(2023YFA1606000), the National Natural Science Foundation of China with Grant No.~12275277, 12447154 and No.~12435004, by the MKW NRW under the funding code NW21-024-A,
by the Deutsche Forschungsgemeinschaft (DFG, German Research Foundation) under Germany's Excellence Strategy -- EXC 3107 -- 533766364
and by the CAS President's International Fellowship Initiative (PIFI) under Grant No.~2025PD0022.

\appendix

\section{Data points from lattice QCD}

\label{app:data}
We list the input data from lattice QCD used in our global fit in Tab. \ref{tab:fbpi_form_factors}.

\begin{table}[htbp]
\centering
\renewcommand{\arraystretch}{1.5}
\caption{$B\to \pi,K,D$ form factors from RBC/UKQCD \cite{Flynn:2015mha}, FNAL/MILC \cite{FermilabLattice:2015mwy,Bailey:2015dka,FermilabLattice:2015ilb} and HPQCD collaborations \cite{Parrott:2022rgu,Na:2015kha}.}
\begin{tabular}{|l|ccc|ccc|}
\hline
 & \multicolumn{3}{c|}{RBC/UKQCD \cite{Flynn:2015mha}} & \multicolumn{3}{c|}{FNAL/MILC \cite{FermilabLattice:2015mwy}} \\
\hline
$q^2$ [GeV$^2$] & 19.0 & 22.6 & 25.1 & 18.0 & 22.0 & 26.0 \\
\hline
$f_{B\pi}^{+}$ & 1.21(16) & 2.27(23) & 4.11(78) & 1.02(6) & 1.98(6) & 6.45(27) \\
$f_{B\pi}^{0}$ & 0.46(7) & 0.68(7) & 0.92(8) & 0.42(3) & - & 0.96(3) \\
\hline  \hline
 & \multicolumn{3}{c|}{FNAL/MILC \cite{Bailey:2015dka}}  & \multicolumn{3}{c|}{HPQCD \cite{Parrott:2022rgu}} \\
\hline
$q^2$ [GeV$^2$] & 19.17 & 21.13 & 22.86 & 19.19 & 21.15 & 22.88 \\
\hline
$f_{BK}^{+}$ & 1.518(36) & 1.993(46) & 2.667(65) & 1.563(49) & 2.067(70) & 2.787(108) \\
$f_{BK}^{0}$ & 0.663(14) & - & 0.852(18) & 0.662(12) & - & 0.847(18) \\
$f_{BK}^{T}$ & 1.529(45) & 2.015(59) & 2.695(82) & 1.532(70) & 2.023(100) & 2.724(151) \\
\hline
\hline
 & \multicolumn{3}{c|}{FNAL/MILC~\cite{FermilabLattice:2015ilb}} & \multicolumn{3}{c|}{HPQCD~\cite{Na:2015kha}} \\
\hline
$q^2$ [GeV$^2$] & 8.51 & 10.08 & 11.66 & 9.30 & 10.48 & 11.46 \\
\hline
$f_{BD}^{+}$ & 1.005(12) & 1.094(10) & 1.199(10) & 1.038(39) & 1.105(42) & 1.167(45) \\
$f_{BD}^{0}$ & 0.825(9) & 0.861(8) & 0.903(7) & 0.840(37) & 0.870(39) & 0.897(40) \\
\hline
\end{tabular}
\label{tab:fbpi_form_factors}
\end{table}

\section{A comparison between LCSRs with $B$-meson LCDAs and LCSRs with light meson LCDAs}
\label{app:ann}

In this section, we determine the parameter $\lambda_B$ by matching the $B$-meson LCSR predictions for the form factors at zero momentum transfer $q^{2}=0$ to given values obtained from different methods. Subsequently, we present the results obtained by incorporating the light meson LCSR results into the global fit.

\begin{figure}[htbp]
    \centering
    \begin{subfigure}[b]{0.5\textwidth}
        \centering
        \includegraphics[width=0.7\textwidth]{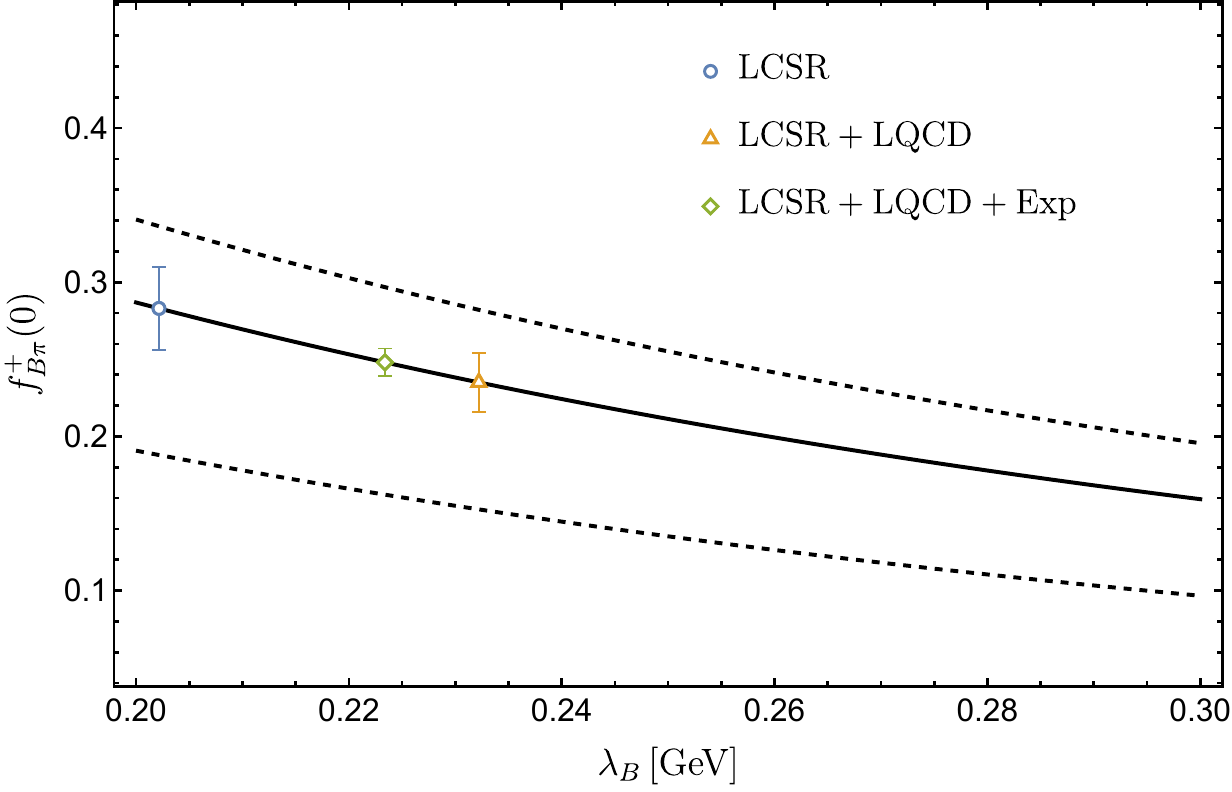}
    \end{subfigure}
    \begin{subfigure}[b]{1\textwidth}
        \centering
        \vspace{10pt}
        \includegraphics[width=0.35\textwidth]{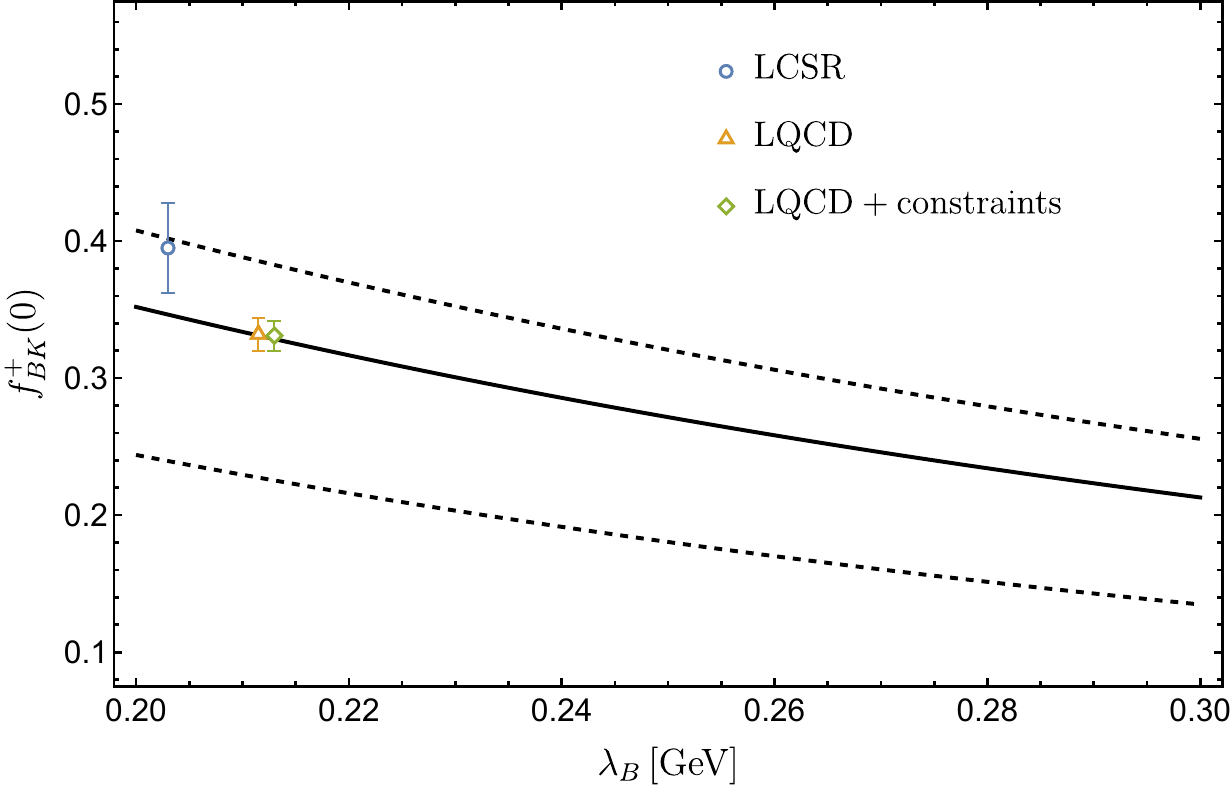}
        \hspace{10pt}
        \includegraphics[width=0.35\textwidth]{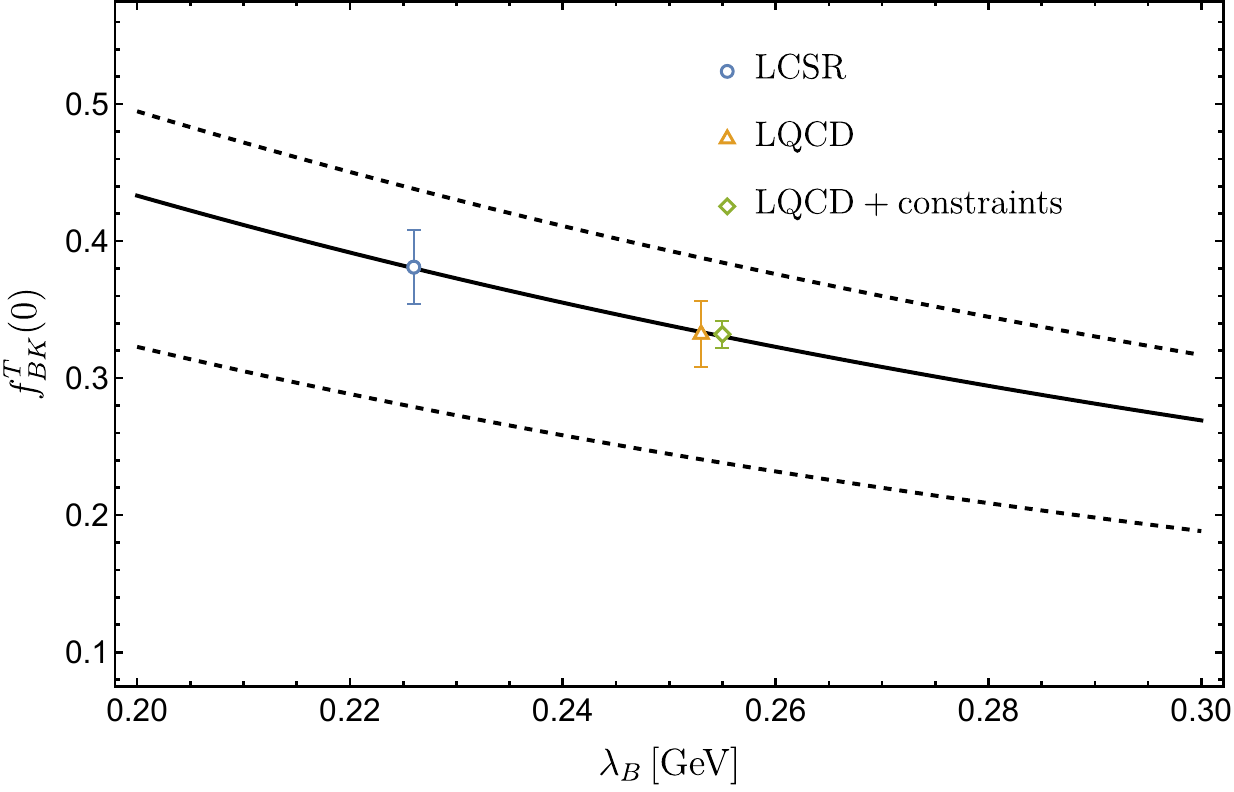}
    \end{subfigure}
    \caption{The black curves show the $\lambda_B$ dependence of the $B\to \pi$ and $B\to K$ form factors predicted by the LCSRs with $B$-meson LCDAs at $q^{2}=0$ (solid curve) with the corresponding uncertainties associated with the LCDA model dependence (dashed curve). The $f_{B\pi}^{+}(0)$ and $f_{BK}^{+,T}(0)$ data points inputs are taken from Refs.~\cite{Khodjamirian:2017fxg,Leljak:2021vte,Gubernari:2023puw}}
    \label{fig:app1}
\end{figure}

By matching the $B$-meson LCSR predictions for $f_{B\pi}^{+}(0)$ and $f_{BK}^{+,T}(0)$ to the given values obtained from various approaches, including the direct light meson LCSR calculations, lattice QCD results, and combined fits with modified BCL parametrization, we determine the values of $\lambda_B$, as shown in Fig.~\ref{fig:app1}. Selected results are summarized below. The first uncertainty arises from the uncertainty of the corresponding form factor input, while the second uncertainty reflects the model dependence of the $B$-meson LCDA.
\begin{equation}
    \begin{aligned}
        & \lambda_B = 223(6)^{+33}_{-23}\,\text{MeV} \,\,\,\, \text{for} \,\,\,\, f_{B\pi}^+(0)=0.248(9),
        \text{\cite{Leljak:2021vte}} \\
           & \lambda_B = 211(6)^{+39}_{-11}\,\text{MeV} \,\,\,\, \text{for} \,\,\,\, f_{BK}^+(0)=0.331(11), 
        \text{\cite{Gubernari:2023puw}} \\
            & \lambda_B = 254(7)^{+44}_{-52}\,\text{MeV} \,\,\,\, \text{for} \,\,\,\, f_{BK}^T(0)=0.330(10). 
        \text{\cite{Gubernari:2023puw}} 
    \end{aligned}
\end{equation}
\begin{figure}[htbp]
    \centering
    \begin{subfigure}[b]{0.5\textwidth}
        \centering
        \includegraphics[width=0.7\textwidth]{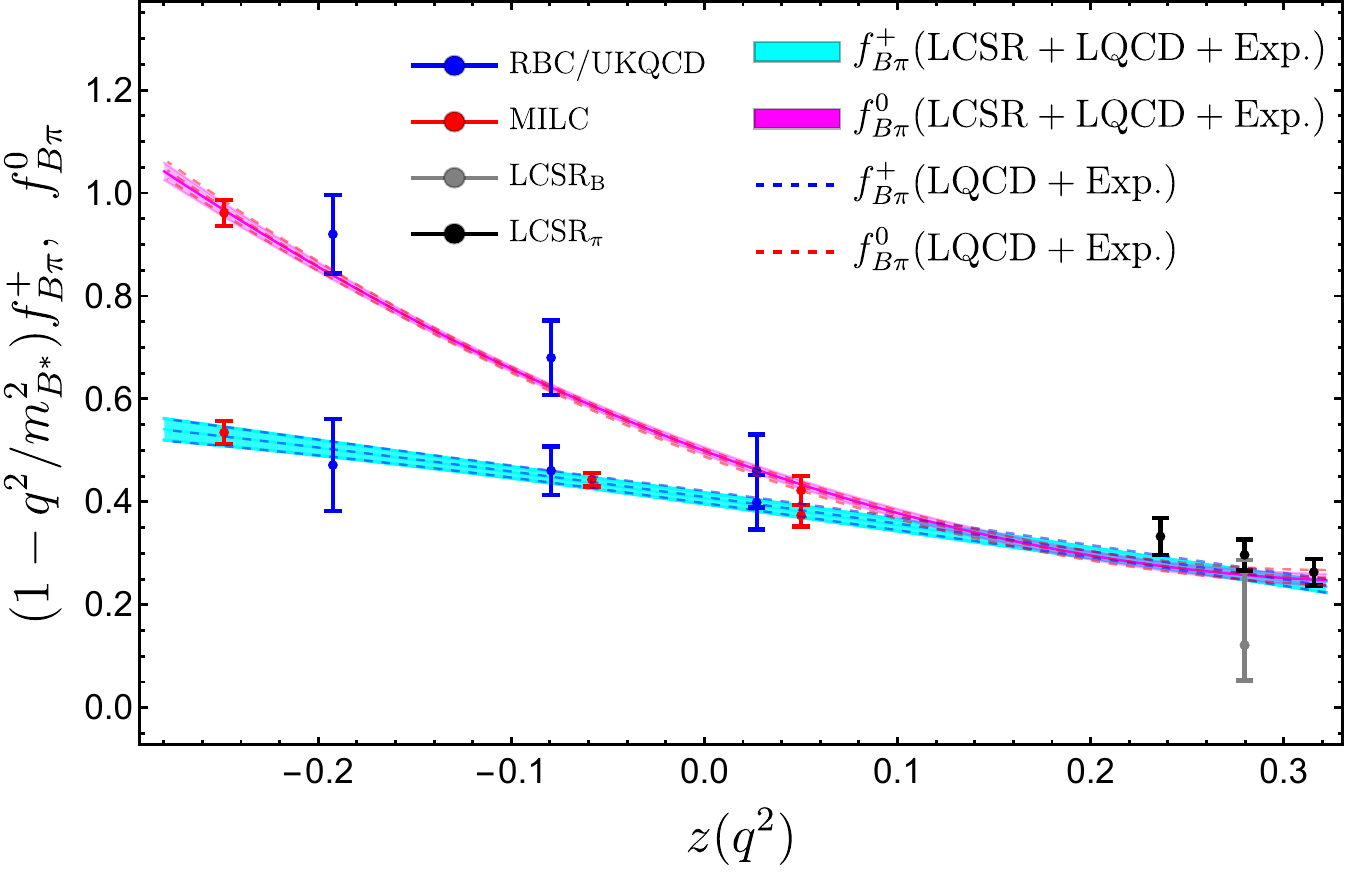}
    \end{subfigure}
    \begin{subfigure}[b]{1\textwidth}
        \centering
        \vspace{10pt}
        \includegraphics[width=0.35\textwidth]{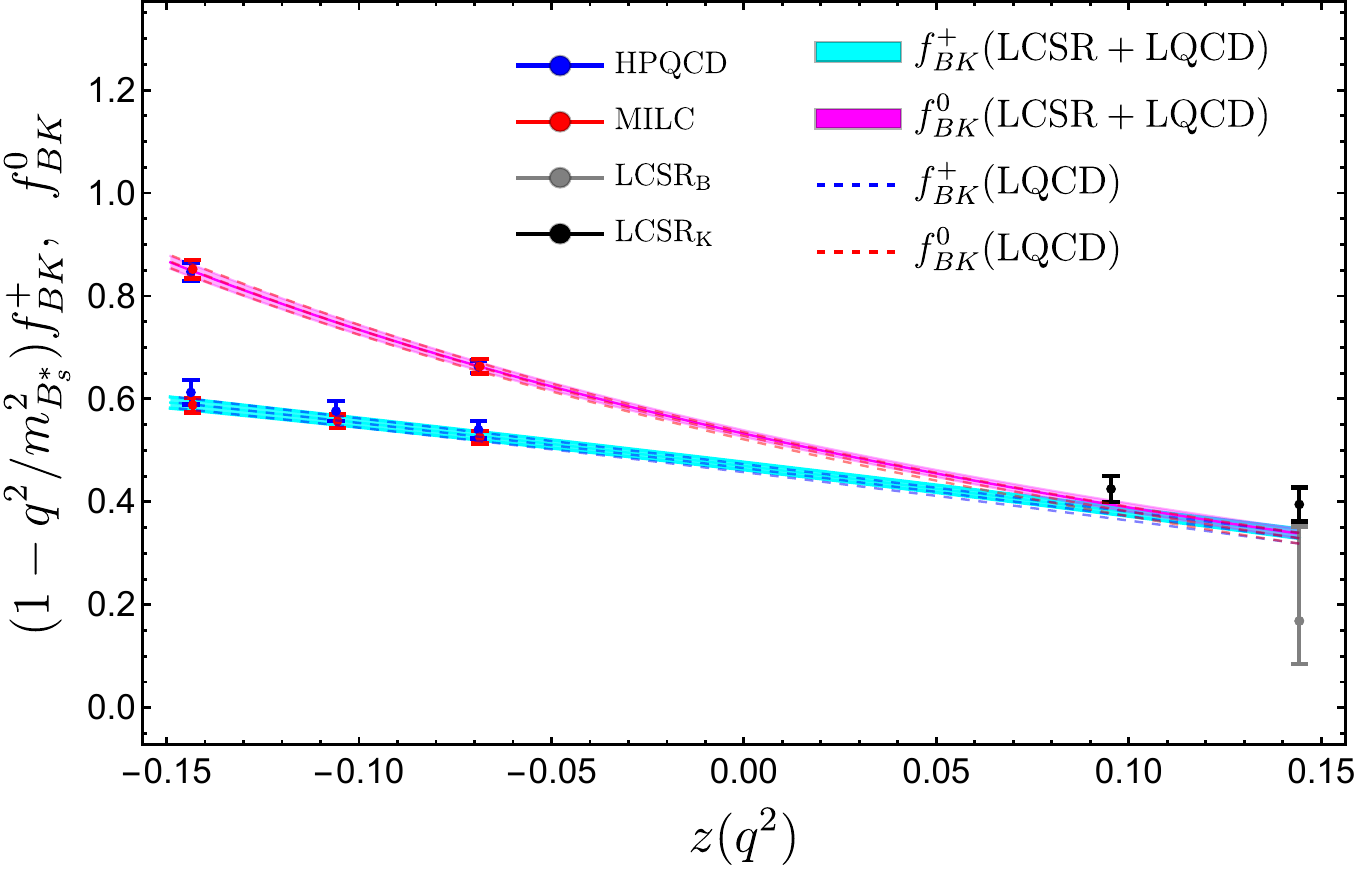}
        \hspace{10pt}
        \includegraphics[width=0.35\textwidth]{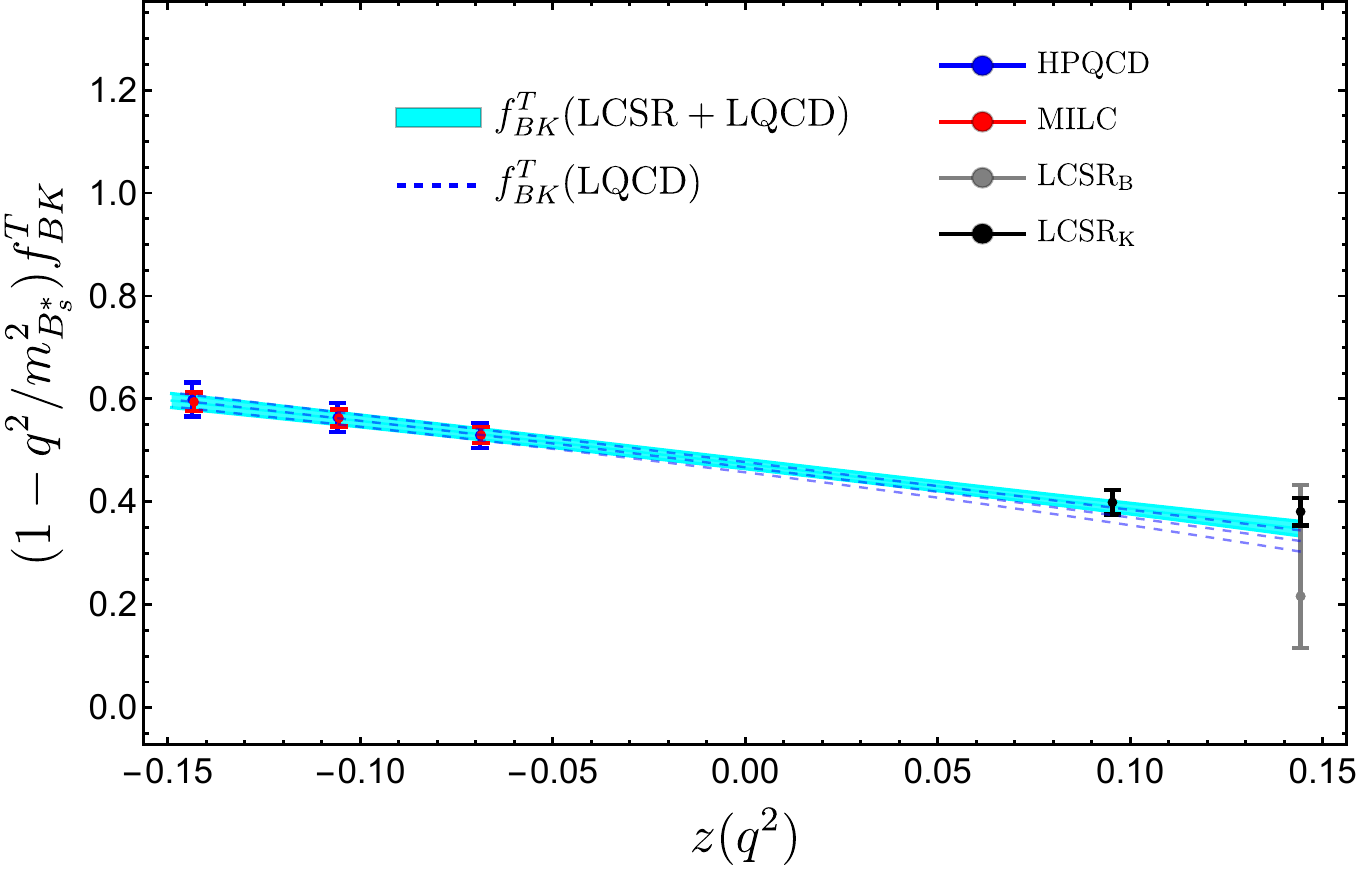}
    \end{subfigure}
    \caption{Results of the global fit to the form factors after including the light meson LCSR inputs as additional constraints (colored bands). The black points (LCSR$_{\pi,K}$) denote the central values and uncertainties of the light meson LCSR inputs. The gray points (LCSR$_B$) show the central values of the $B$-meson LCSR predictions for $\lambda_B=350~\mathrm{MeV}$ with the error bars corresponding to $\lambda_B=200~\mathrm{MeV}$ and $500~\mathrm{MeV}$, respectively.}
    \label{fig:app2}
\end{figure}
After incorporating the correlated light meson LCSR inputs for $f_{B\pi}^{+}$ and $f_{BK}^{+,T}$, the global fit yields $f_{B\pi}^{+}(0)=0.258(10)$, $f_{BK}^{+}(0)=0.338(10)$, $f_{BK}^{T}(0)=0.348(17)$ and $\lambda_B=215(19)~\mathrm{MeV}$, as shown in Fig.~\ref{fig:app2}. These results are consistent with those obtained from other methods, such as $\{f_{B\pi}^{+}(0),f_{BK}^{+}(0),f_{BK}^{T}(0)\}=\{0.259(12),0.330(10),
\notag
\\
0.335(19)\}$ from Eq. (\ref{chi2tot}) and Refs. \cite{Leljak:2021vte,Gubernari:2023puw}.

This consistent can be understood as follows. For the $B\to\pi$ channel, we include the experimentally measured branching fractions, which effectively constrain the extrapolation uncertainty. For the $B\to K$ channel, the precise HPQCD 2022 lattice results provide direct access to $f_{BK}(q^{2})$ down to $q^{2}=0$. Consequently, in our analysis we use the light meson LCSR inputs for cross checks.


\begin{thebibliography}{99}
\bibitem{Lepage:1979zb}
G.~P.~Lepage and S.~J.~Brodsky,
Phys. Lett. B \textbf{87} (1979), 359-365
doi:10.1016/0370-2693(79)90554-9

\bibitem{Efremov:1979qk}
A.~V.~Efremov and A.~V.~Radyushkin,
Phys. Lett. B \textbf{94} (1980), 245-250
doi:10.1016/0370-2693(80)90869-2

\bibitem{Grozin:1996pq}
A.~G.~Grozin and M.~Neubert,
Phys. Rev. D \textbf{55} (1997), 272-290
doi:10.1103/PhysRevD.55.272
[arXiv:hep-ph/9607366 [hep-ph]].

\bibitem{Beneke:1999br}
M.~Beneke, G.~Buchalla, M.~Neubert and C.~T.~Sachrajda,
Phys. Rev. Lett. \textbf{83} (1999), 1914-1917
doi:10.1103/PhysRevLett.83.1914
[arXiv:hep-ph/9905312 [hep-ph]].

\bibitem{Keum:2003js}
Y.~Y.~Keum, T.~Kurimoto, H.~N.~Li, C.~D.~Lu and A.~I.~Sanda,
Phys. Rev. D \textbf{69} (2004), 094018
doi:10.1103/PhysRevD.69.094018
[arXiv:hep-ph/0305335 [hep-ph]].

\bibitem{Hua:2020gnw}
J.~Hua \textit{et al.} [Lattice Parton],
Phys. Rev. Lett. \textbf{127} (2021) no.6, 062002
doi:10.1103/PhysRevLett.127.062002
[arXiv:2011.09788 [hep-lat]].

\bibitem{LatticeParton:2022zqc}
J.~Hua \textit{et al.} [Lattice Parton],
Phys. Rev. Lett. \textbf{129} (2022) no.13, 132001
doi:10.1103/PhysRevLett.129.132001
[arXiv:2201.09173 [hep-lat]].

\bibitem{Khodjamirian:2005ea}
A.~Khodjamirian, T.~Mannel and N.~Offen,
Phys. Lett. B \textbf{620} (2005), 52-60
doi:10.1016/j.physletb.2005.06.021
[arXiv:hep-ph/0504091 [hep-ph]].

\bibitem{Braun:2012kp}
V.~M.~Braun and A.~Khodjamirian,
Phys. Lett. B \textbf{718} (2013), 1014-1019
doi:10.1016/j.physletb.2012.11.047
[arXiv:1210.4453 [hep-ph]].

\bibitem{Cheng:2014fwa}
S.~Cheng, Y.~Y.~Fan, X.~Yu, C.~D.~L{\"u} and Z.~J.~Xiao,
Phys. Rev. D \textbf{89} (2014) no.9, 094004
doi:10.1103/PhysRevD.89.094004
[arXiv:1402.5501 [hep-ph]].

\bibitem{Wang:2015vgv}
Y.~M.~Wang and Y.~L.~Shen,
Nucl. Phys. B \textbf{898} (2015), 563-604
doi:10.1016/j.nuclphysb.2015.07.016
[arXiv:1506.00667 [hep-ph]].

\bibitem{Beneke:2020fot}
M.~Beneke, C.~Bobeth and Y.~M.~Wang,
JHEP \textbf{12} (2020), 148
doi:10.1007/JHEP12(2020)148
[arXiv:2008.12494 [hep-ph]].

\bibitem{Cui:2023jiw}
B.~Y.~Cui, Y.~K.~Huang, Y.~M.~Wang and X.~C.~Zhao,
Phys. Rev. D \textbf{108} (2023) no.7, L071504
doi:10.1103/PhysRevD.108.L071504
[arXiv:2301.12391 [hep-ph]].

\bibitem{Shen:2020hfq}
Y.~L.~Shen, Y.~M.~Wang and Y.~B.~Wei,
JHEP \textbf{12} (2020), 169
doi:10.1007/JHEP12(2020)169
[arXiv:2009.02723 [hep-ph]].

\bibitem{Gao:2024vql}
J.~Gao, U.~G.~Mei{\ss}ner, Y.~L.~Shen and D.~H.~Li,
Phys. Rev. D \textbf{112} (2025) no.1, 1
doi:10.1103/yvjd-2ymn
[arXiv:2412.13084 [hep-ph]].

\bibitem{Huang:2024xii}
Y.~K.~Huang, Y.~L.~Shen, C.~Wang and Y.~M.~Wang,
Phys. Rev. Lett. \textbf{134} (2025) no.9, 091901
doi:10.1103/PhysRevLett.134.091901
[arXiv:2403.11258 [hep-ph]].

\bibitem{Lange:2003ff}
B.~O.~Lange and M.~Neubert,
Phys. Rev. Lett. \textbf{91} (2003), 102001
doi:10.1103/PhysRevLett.91.102001
[arXiv:hep-ph/0303082 [hep-ph]].

\bibitem{Bell:2013tfa}
G.~Bell, T.~Feldmann, Y.~M.~Wang and M.~W.~Y.~Yip,
JHEP \textbf{11} (2013), 191
doi:10.1007/JHEP11(2013)191
[arXiv:1308.6114 [hep-ph]].

\bibitem{Braun:2014owa}
V.~M.~Braun and A.~N.~Manashov,
Phys. Lett. B \textbf{731} (2014), 316-319
doi:10.1016/j.physletb.2014.02.051
[arXiv:1402.5822 [hep-ph]].

\bibitem{Galda:2020epp}
A.~M.~Galda and M.~Neubert,
Phys. Rev. D \textbf{102} (2020), 071501
doi:10.1103/PhysRevD.102.071501
[arXiv:2006.05428 [hep-ph]].

\bibitem{Huang:2023jdu}
Y.~K.~Huang, Y.~Ji, Y.~L.~Shen, C.~Wang, Y.~M.~Wang and X.~C.~Zhao,
Phys. Rev. Lett. \textbf{133} (2024) no.17, 171901
doi:10.1103/PhysRevLett.133.171901
[arXiv:2312.15439 [hep-ph]].

\bibitem{Feldmann:2022uok}
T.~Feldmann, P.~L{\"u}ghausen and D.~van Dyk,
JHEP \textbf{10} (2022), 162
doi:10.1007/JHEP10(2022)162
[arXiv:2203.15679 [hep-ph]].

\bibitem{Lee:2005gza}
S.~J.~Lee and M.~Neubert,
Phys. Rev. D \textbf{72} (2005), 094028
doi:10.1103/PhysRevD.72.094028
[arXiv:hep-ph/0509350 [hep-ph]].

\bibitem{Feldmann:2014ika}
T.~Feldmann, B.~O.~Lange and Y.~M.~Wang,
Phys. Rev. D \textbf{89} (2014) no.11, 114001
doi:10.1103/PhysRevD.89.114001
[arXiv:1404.1343 [hep-ph]].

\bibitem{Beneke:2023nmj}
M.~Beneke, G.~Finauri, K.~K.~Vos and Y.~Wei,
JHEP \textbf{09} (2023), 066
doi:10.1007/JHEP09(2023)066
[arXiv:2305.06401 [hep-ph]].

\bibitem{Wang:2019tgg}
W.~Wang, J.~H.~Zhang, S.~Zhao and R.~Zhu,
Phys. Rev. D \textbf{100} (2019) no.7, 074509
doi:10.1103/PhysRevD.100.074509
[arXiv:1904.00978 [hep-ph]].

\bibitem{Wang:2019msf}
W.~Wang, Y.~M.~Wang, J.~Xu and S.~Zhao,
Phys. Rev. D \textbf{102} (2020) no.1, 011502
doi:10.1103/PhysRevD.102.011502
[arXiv:1908.09933 [hep-ph]].

\bibitem{Wang:2024wwa}
W.~Wang, J.~Xu, Q.~A.~Zhang and S.~Zhao,
[arXiv:2411.07101 [hep-ph]].

\bibitem{LatticeParton:2024zko}
X.~Y.~Han \textit{et al.} [Lattice Parton],
Phys. Rev. D \textbf{111} (2025) no.3, 034503
doi:10.1103/PhysRevD.111.034503
[arXiv:2410.18654 [hep-lat]].

\bibitem{Braun:2017liq}
V.~M.~Braun, Y.~Ji and A.~N.~Manashov,
JHEP \textbf{05} (2017), 022
doi:10.1007/JHEP05(2017)022
[arXiv:1703.02446 [hep-ph]].

\bibitem{Beneke:2018wjp}
M.~Beneke, V.~M.~Braun, Y.~Ji and Y.~B.~Wei,
JHEP \textbf{07} (2018), 154
doi:10.1007/JHEP07(2018)154
[arXiv:1804.04962 [hep-ph]].

\bibitem{Braun:2003wx}
V.~M.~Braun, D.~Y.~Ivanov and G.~P.~Korchemsky,
Phys. Rev. D \textbf{69} (2004), 034014
doi:10.1103/PhysRevD.69.034014
[arXiv:hep-ph/0309330 [hep-ph]].

\bibitem{Khodjamirian:2020hob}
A.~Khodjamirian, R.~Mandal and T.~Mannel,
JHEP \textbf{10} (2020), 043
doi:10.1007/JHEP10(2020)043
[arXiv:2008.03935 [hep-ph]].

\bibitem{Belle:2018jqd}
M.~Gelb \textit{et al.} [Belle],
Phys. Rev. D \textbf{98} (2018) no.11, 112016
doi:10.1103/PhysRevD.98.112016
[arXiv:1810.12976 [hep-ex]].

\bibitem{Wang:2016qii}
Y.~M.~Wang,
JHEP \textbf{09} (2016), 159
doi:10.1007/JHEP09(2016)159
[arXiv:1606.03080 [hep-ph]].

\bibitem{Han:2024fkr}
X.~Y.~Han, J.~Hua, X.~Ji, C.~D.~L{\"u}, W.~Wang, J.~Xu, Q.~A.~Zhang and S.~Zhao,
Phys. Rev. D \textbf{111} (2025) no.11, L111503
doi:10.1103/2t8s-w8t6
[arXiv:2403.17492 [hep-ph]].

\bibitem{Janowski:2021yvz}
T.~Janowski, B.~Pullin and R.~Zwicky,
JHEP \textbf{12} (2021), 008
doi:10.1007/JHEP12(2021)008
[arXiv:2106.13616 [hep-ph]].

\bibitem{Gao:2019lta}
J.~Gao, C.~D.~L{\"u}, Y.~L.~Shen, Y.~M.~Wang and Y.~B.~Wei,
Phys. Rev. D \textbf{101} (2020) no.7, 074035
doi:10.1103/PhysRevD.101.074035
[arXiv:1907.11092 [hep-ph]].

\bibitem{Mandal:2023lhp}
R.~Mandal, S.~Nandi and I.~Ray,
Phys. Lett. B \textbf{848} (2024), 138345
doi:10.1016/j.physletb.2023.138345
[arXiv:2308.07033 [hep-ph]].

\bibitem{Flynn:2015mha}
J.~M.~Flynn, T.~Izubuchi, T.~Kawanai, C.~Lehner, A.~Soni, R.~S.~Van de Water and O.~Witzel,
Phys. Rev. D \textbf{91} (2015) no.7, 074510
doi:10.1103/PhysRevD.91.074510
[arXiv:1501.05373 [hep-lat]].

\bibitem{FermilabLattice:2015mwy}
J.~A.~Bailey \textit{et al.} [Fermilab Lattice and MILC],
Phys. Rev. D \textbf{92} (2015) no.1, 014024
doi:10.1103/PhysRevD.92.014024
[arXiv:1503.07839 [hep-lat]].

\bibitem{Bailey:2015dka}
J.~A.~Bailey, A.~Bazavov, C.~Bernard, C.~M.~Bouchard, C.~DeTar, D.~Du, A.~X.~El-Khadra, J.~Foley, E.~D.~Freeland and E.~G{\'a}miz, \textit{et al.}
Phys. Rev. D \textbf{93} (2016) no.2, 025026
doi:10.1103/PhysRevD.93.025026
[arXiv:1509.06235 [hep-lat]].

\bibitem{FermilabLattice:2015ilb}
J.~A.~Bailey \textit{et al.} [Fermilab Lattice and MILC],
Phys. Rev. D \textbf{92} (2015) no.3, 034506
doi:10.1103/PhysRevD.92.034506
[arXiv:1503.07237 [hep-lat]].

\bibitem{Parrott:2022rgu}
W.~G.~Parrott \textit{et al.} [(HPQCD collaboration){\textsection} and HPQCD],
Phys. Rev. D \textbf{107} (2023) no.1, 014510
doi:10.1103/PhysRevD.107.014510
[arXiv:2207.12468 [hep-lat]].

\bibitem{Na:2015kha}
H.~Na \textit{et al.} [HPQCD],
Phys. Rev. D \textbf{92} (2015) no.5, 054510
[erratum: Phys. Rev. D \textbf{93} (2016) no.11, 119906]
doi:10.1103/PhysRevD.93.119906
[arXiv:1505.03925 [hep-lat]].

\bibitem{BaBar:2010efp}
P.~del Amo Sanchez \textit{et al.} [BaBar],
Phys. Rev. D \textbf{83} (2011), 032007
doi:10.1103/PhysRevD.83.032007
[arXiv:1005.3288 [hep-ex]].

\bibitem{BaBar:2012thb}
J.~P.~Lees \textit{et al.} [BaBar],
Phys. Rev. D \textbf{86} (2012), 092004
doi:10.1103/PhysRevD.86.092004
[arXiv:1208.1253 [hep-ex]].

\bibitem{Belle:2010hep}
H.~Ha \textit{et al.} [Belle],
Phys. Rev. D \textbf{83} (2011), 071101
doi:10.1103/PhysRevD.83.071101
[arXiv:1012.0090 [hep-ex]].

\bibitem{Belle:2013hlo}
A.~Sibidanov \textit{et al.} [Belle],
Phys. Rev. D \textbf{88} (2013) no.3, 032005
doi:10.1103/PhysRevD.88.032005
[arXiv:1306.2781 [hep-ex]].

\bibitem{Belle-II:2022imn}
K.~Adamczyk \textit{et al.} [Belle-II],
[arXiv:2210.04224 [hep-ex]].

\bibitem{Cui:2022zwm}
B.~Y.~Cui, Y.~K.~Huang, Y.~L.~Shen, C.~Wang and Y.~M.~Wang,
JHEP \textbf{03} (2023), 140
doi:10.1007/JHEP03(2023)140
[arXiv:2212.11624 [hep-ph]].

\bibitem{Gao:2021sav}
J.~Gao, T.~Huber, Y.~Ji, C.~Wang, Y.~M.~Wang and Y.~B.~Wei,
JHEP \textbf{05} (2022), 024
doi:10.1007/JHEP05(2022)024
[arXiv:2112.12674 [hep-ph]].

\bibitem{Balitsky:1986st}
I.~I.~Balitsky, V.~M.~Braun and A.~V.~Kolesnichenko,
Sov. J. Nucl. Phys. \textbf{44} (1986), 1028

\bibitem{Balitsky:1989ry}
I.~I.~Balitsky, V.~M.~Braun and A.~V.~Kolesnichenko,
Nucl. Phys. B \textbf{312} (1989), 509-550
doi:10.1016/0550-3213(89)90570-1

\bibitem{Chernyak:1990ag}
V.~L.~Chernyak and I.~R.~Zhitnitsky,
Nucl. Phys. B \textbf{345} (1990), 137-172
doi:10.1016/0550-3213(90)90612-H

\bibitem{hep-ph/9305348}
V.~M.~Belyaev, A.~Khodjamirian and R.~Ruckl,
Z. Phys. C \textbf{60} (1993), 349-356
doi:10.1007/BF01474633
[arXiv:hep-ph/9305348 [hep-ph]].

\bibitem{hep-ph/9701238}
P.~Ball and V.~M.~Braun,
Phys. Rev. D \textbf{55} (1997), 5561-5576
doi:10.1103/PhysRevD.55.5561
[arXiv:hep-ph/9701238 [hep-ph]].

\bibitem{Khodjamirian:2006st}
A.~Khodjamirian, T.~Mannel and N.~Offen,
Phys. Rev. D \textbf{75} (2007), 054013
doi:10.1103/PhysRevD.75.054013
[arXiv:hep-ph/0611193 [hep-ph]].

\bibitem{Khodjamirian:2023wol}
A.~Khodjamirian, B.~Meli{\'c} and Y.~M.~Wang,
Eur. Phys. J. ST \textbf{233} (2024) no.2, 271-298
doi:10.1140/epjs/s11734-023-01046-6
[arXiv:2311.08700 [hep-ph]].

\bibitem{Wang:2017jow}
Y.~M.~Wang, Y.~B.~Wei, Y.~L.~Shen and C.~D.~L{\"u},
JHEP \textbf{06} (2017), 062
doi:10.1007/JHEP06(2017)062
[arXiv:1701.06810 [hep-ph]].

\bibitem{Gubernari:2018wyi}
N.~Gubernari, A.~Kokulu and D.~van Dyk,
JHEP \textbf{01} (2019), 150
doi:10.1007/JHEP01(2019)150
[arXiv:1811.00983 [hep-ph]].

\bibitem{Wang:2018wfj}
Y.~M.~Wang and Y.~L.~Shen,
JHEP \textbf{05} (2018), 184
doi:10.1007/JHEP05(2018)184
[arXiv:1803.06667 [hep-ph]].

\bibitem{FlavourLatticeAveragingGroupFLAG:2024oxs}
Y.~Aoki \textit{et al.} [Flavour Lattice Averaging Group (FLAG)],
Phys. Rev. D \textbf{113} (2026) no.1, 014508
doi:10.1103/nfzp-p5dn
[arXiv:2411.04268 [hep-lat]].

\bibitem{ParticleDataGroup:2024cfk}
S.~Navas \textit{et al.} [Particle Data Group],
Phys. Rev. D \textbf{110} (2024) no.3, 030001
doi:10.1103/PhysRevD.110.030001

\bibitem{Dowdall:2012ab}
R.~J.~Dowdall, C.~T.~H.~Davies, T.~C.~Hammant and R.~R.~Horgan,
Phys. Rev. D \textbf{86} (2012), 094510
doi:10.1103/PhysRevD.86.094510
[arXiv:1207.5149 [hep-lat]].

\bibitem{Mathur:2018epb}
N.~Mathur, M.~Padmanath and S.~Mondal,
Phys. Rev. Lett. \textbf{121} (2018) no.20, 202002
doi:10.1103/PhysRevLett.121.202002
[arXiv:1806.04151 [hep-lat]].

\bibitem{Bardeen:2003kt}
W.~A.~Bardeen, E.~J.~Eichten and C.~T.~Hill,
Phys. Rev. D \textbf{68} (2003), 054024
doi:10.1103/PhysRevD.68.054024
[arXiv:hep-ph/0305049 [hep-ph]].

\bibitem{Bourrely:2008za}
C.~Bourrely, I.~Caprini and L.~Lellouch,
Phys. Rev. D \textbf{79} (2009), 013008
[erratum: Phys. Rev. D \textbf{82} (2010), 099902]
doi:10.1103/PhysRevD.82.099902
[arXiv:0807.2722 [hep-ph]].

\bibitem{Boyd:1994tt}
C.~G.~Boyd, B.~Grinstein and R.~F.~Lebed,
Phys. Rev. Lett. \textbf{74} (1995), 4603-4606
doi:10.1103/PhysRevLett.74.4603
[arXiv:hep-ph/9412324 [hep-ph]].

\bibitem{Boyd:1995sq}
C.~G.~Boyd, B.~Grinstein and R.~F.~Lebed,
Nucl. Phys. B \textbf{461} (1996), 493-511
doi:10.1016/0550-3213(95)00653-2
[arXiv:hep-ph/9508211 [hep-ph]].

\bibitem{Boyd:1997kz}
C.~G.~Boyd, B.~Grinstein and R.~F.~Lebed,
Phys. Rev. D \textbf{56} (1997), 6895-6911
doi:10.1103/PhysRevD.56.6895
[arXiv:hep-ph/9705252 [hep-ph]].

\bibitem{Gambino:2019sif}
P.~Gambino, M.~Jung and S.~Schacht,
Phys. Lett. B \textbf{795} (2019), 386-390
doi:10.1016/j.physletb.2019.06.039
[arXiv:1905.08209 [hep-ph]].

\bibitem{DiCarlo:2021dzg}
M.~Di Carlo, G.~Martinelli, M.~Naviglio, F.~Sanfilippo, S.~Simula and L.~Vittorio,
Phys. Rev. D \textbf{104} (2021) no.5, 054502
doi:10.1103/PhysRevD.104.054502
[arXiv:2105.02497 [hep-lat]].

\bibitem{Martinelli:2021onb}
G.~Martinelli, S.~Simula and L.~Vittorio,
Phys. Rev. D \textbf{105} (2022) no.3, 034503
doi:10.1103/PhysRevD.105.034503
[arXiv:2105.08674 [hep-ph]].

\bibitem{Jaiswal:2020wer}
S.~Jaiswal, S.~Nandi and S.~K.~Patra,
JHEP \textbf{06} (2020), 165
doi:10.1007/JHEP06(2020)165
[arXiv:2002.05726 [hep-ph]].

\bibitem{Leljak:2021vte}
D.~Leljak, B.~Meli{\'c} and D.~van Dyk,
JHEP \textbf{07} (2021), 036
doi:10.1007/JHEP07(2021)036
[arXiv:2102.07233 [hep-ph]].

\bibitem{Lellouch:1995yv}
L.~Lellouch,
Nucl. Phys. B \textbf{479} (1996), 353-391
doi:10.1016/0550-3213(96)00443-9
[arXiv:hep-ph/9509358 [hep-ph]].

\bibitem{Becher:2005bg}
T.~Becher and R.~J.~Hill,
Phys. Lett. B \textbf{633} (2006), 61-69
doi:10.1016/j.physletb.2005.11.063
[arXiv:hep-ph/0509090 [hep-ph]].

\bibitem{Duplancic:2008ix}
G.~Duplancic, A.~Khodjamirian, T.~Mannel, B.~Melic and N.~Offen,
JHEP \textbf{04} (2008), 014
doi:10.1088/1126-6708/2008/04/014
[arXiv:0801.1796 [hep-ph]].

\bibitem{Bharucha:2010im}
A.~Bharucha, T.~Feldmann and M.~Wick,
JHEP \textbf{09} (2010), 090
doi:10.1007/JHEP09(2010)090
[arXiv:1004.3249 [hep-ph]].

\bibitem{Bharucha:2012wy}
A.~Bharucha,
JHEP \textbf{05} (2012), 092
doi:10.1007/JHEP05(2012)092
[arXiv:1203.1359 [hep-ph]].

\bibitem{Khodjamirian:2011ub}
A.~Khodjamirian, T.~Mannel, N.~Offen and Y.~M.~Wang,
Phys. Rev. D \textbf{83} (2011), 094031
doi:10.1103/PhysRevD.83.094031
[arXiv:1103.2655 [hep-ph]].

\bibitem{SentitemsuImsong:2014plu}
I.~Sentitemsu Imsong, A.~Khodjamirian, T.~Mannel and D.~van Dyk,
JHEP \textbf{02} (2015), 126
doi:10.1007/JHEP02(2015)126
[arXiv:1409.7816 [hep-ph]].

\bibitem{Khodjamirian:2017fxg}
A.~Khodjamirian and A.~V.~Rusov,
JHEP \textbf{08} (2017), 112
doi:10.1007/JHEP08(2017)112
[arXiv:1703.04765 [hep-ph]].

\bibitem{Gonzalez-Solis:2021pyh}
S.~Gonz{\`a}lez-Sol{\'\i}s, P.~Masjuan and C.~Rojas,
Phys. Rev. D \textbf{104} (2021) no.11, 114041
doi:10.1103/PhysRevD.104.114041
[arXiv:2110.06153 [hep-ph]].

\bibitem{Gambino:2020jvv}
P.~Gambino, A.~S.~Kronfeld, M.~Rotondo, C.~Schwanda, F.~Bernlochner, A.~Bharucha, C.~Bozzi, M.~Calvi, L.~Cao and G.~Ciezarek, \textit{et al.}
Eur. Phys. J. C \textbf{80} (2020) no.10, 966
doi:10.1140/epjc/s10052-020-08490-x
[arXiv:2006.07287 [hep-ph]].

\bibitem{EOSAuthors:2021xpv}
D.~van Dyk \textit{et al.} [EOS Authors],
Eur. Phys. J. C \textbf{82} (2022) no.6, 569
doi:10.1140/epjc/s10052-022-10177-4
[arXiv:2111.15428 [hep-ph]].

\bibitem{Flynn:2023qmi}
J.~M.~Flynn, A.~J{\"u}ttner and J.~T.~Tsang,
JHEP \textbf{12} (2023), 175
doi:10.1007/JHEP12(2023)175
[arXiv:2303.11285 [hep-ph]].

\bibitem{Gubernari:2023puw}
N.~Gubernari, M.~Reboud, D.~van Dyk and J.~Virto,
JHEP \textbf{12} (2023), 153
[erratum: JHEP \textbf{01} (2025), 125]
doi:10.1007/JHEP12(2023)153
[arXiv:2305.06301 [hep-ph]].

\bibitem{Martinelli:2021frl}
G.~Martinelli, S.~Simula and L.~Vittorio,
Phys. Rev. D \textbf{104} (2021) no.9, 094512
doi:10.1103/PhysRevD.104.094512
[arXiv:2105.07851 [hep-lat]].
\end{thebibliography}
\end{document}